\documentclass[aps,prd,twocolumn,showpacs,superscriptaddress,preprintnumbers,notitlepage,nofootinbib]{revtex4-1}
\usepackage[colorlinks=true, pdfstartview=FitV, linkcolor=red,citecolor=blue, urlcolor=blue,
bookmarks=true, bookmarksnumbered=true, breaklinks]{hyperref}
\usepackage[dvipdfmx]{graphicx}

\usepackage{amsmath,amssymb,tabularx,ulem,mathrsfs,bm,multirow}
\usepackage{braket}

\newcommand*{\e}[1]{
\begin{eqnarray}
#1
\end{eqnarray}
}

\def\simge{\mathrel{%
   \rlap{\raise 0.511ex \hbox{$>$}}{\lower 0.511ex \hbox{$\sim$}}}}
\def\simle{\mathrel{
   \rlap{\raise 0.511ex \hbox{$<$}}{\lower 0.511ex \hbox{$\sim$}}}}
\def\bigs{\mathrel{
   \rlap{\raise 0.531ex \hbox{$>$}}{\lower 0.531ex \hbox{$<$}}}}

\begin{document}

\title{Quarkonium radiative decays from the hadronic Paschen-Back effect}

\author{Sachio Iwasaki}
\email{iwasaki.s.aa@m.titech.ac.jp}
\affiliation{Department of Physics, Tokyo Institute of Technology, Meguro, Tokyo, 152-8551, Japan}
\affiliation{Advanced Science Research Center, Japan Atomic Energy Agency, Tokai, Ibaraki, 319-1195, Japan}
\author{Kei Suzuki}
\email{kei.suzuki@kek.jp}
\affiliation{KEK Theory Center, Institute of Particle and Nuclear Studies, High Energy Accelerator Research Organization, 1-1 Oho, Tsukuba, Ibaraki, 305-0801, Japan}

\date{\today}

\preprint{KEK-TH-2056}

\begin{abstract}
We study the radiative (E1 and M1) decays of P-wave quarkonia in a strong magnetic field based on the Lagrangian of potential nonrelativistic QCD.
To investigate their properties, we implement a polarized wave function basis justified in the Paschen-Back limit.
In a magnetic field stronger than the spin-orbit coupling, the wave functions of the P-wave quarkonia are drastically deformed by the Hadronic Paschen-Back effect.
Such deformation leads to the anisotropy of the direction of decays from the P-wave quarkonia.
The analytic formulas for the radiative decay widths in the nonrelativistic limit are shown, and the qualitative decay properties are discussed.
\end{abstract}
\maketitle

\section{Introduction}
The Paschen-Back effect (PBE) is one of the well-known phenomena in atomic physics \cite{Paschen1921}.
When a magnetic field stronger than the scale of the spin-orbit (LS) coupling of a system is applied, namely in the Paschen-Back (PB) region, the eigenstates of the system are approximately characterized only by $L_z$ and $S_z$, where $L_z$ and $S_z$ are the $z$ components (parallel to the magnetic field) of the orbital and spin angular momenta, respectively.
A similar effect occurs even in hadronic systems with finite orbital angular momenta \cite{Iwasaki:2018pby}, which we call the {\it Hadronic Paschen-Back effect} (HPBE).

In Ref.~\cite{Iwasaki:2018pby}, we showed that the wave functions of P-wave charmonia ($h_c$, $\chi_{c0}$, $\chi_{c1}$, and $\chi_{c2}$) in a magnetic field are drastically deformed by the HPBE, and we suggested possible experimental observables of strong magnetic fields.
Such a strong magnetic field is expected to be created in heavy-ion collisions at the Super Proton Synchrotron (SPS), the Relativistic Heavy Ion Collider (RHIC), and the Large Hadron Collider (LHC) \cite{Kharzeev:2007jp,Skokov:2009qp,Voronyuk:2011jd,Ou:2011fm,Bzdak:2011yy,Deng:2012pc,Bloczynski:2012en,Bloczynski:2013mca,Deng:2014uja,Huang:2015oca,Hattori:2016emy,Zhao:2017rpf}, where the maxima of the strength reach $|eB| \sim 0.01$, $0.1$, and $\sim 1.0 \ \mathrm{GeV}^2$, respectively (see Ref.~\cite{Skokov:2009qp} for SPS energy).
Under such environments, the HPBE for the P-wave quarkonia can be related to the observables for properties of the S-wave quarkonia \cite{Marasinghe:2011bt,Tuchin:2011cg,Yang:2011cz,Tuchin:2013ie,Machado:2013rta,Alford:2013jva,Cho:2014exa,Dudal:2014jfa,Cho:2014loa,Guo:2015nsa,Bonati:2015dka,Sadofyev:2015hxa,Suzuki:2016kcs,Yoshida:2016xgm,Suzuki:2016fof,Hoelck:2017dby,Braga:2018zlu}, heavy-light mesons \cite{Machado:2013rta,Machado:2013yaa,Gubler:2015qok,Yoshida:2016xgm,Reddy:2017pqp,Dhale:2018plh}, and heavy-quark diffusion \cite{Fukushima:2015wck,Finazzo:2016mhm,Das:2016cwd,Dudal:2018rki,Chatterjee:2018lsx}\footnote{In fact, experimentally realistic situations are more complex, and various backgrounds (as well as magnetic fields) should be involved in observables.
The transport properties of heavy quarks under a magnetic field might be affected by the chiral drag force in a medium with a chirality imbalance \cite{Rajagopal:2015roa,Stephanov:2015roa, Sadofyev:2015tmb} and/or the magnetically induced QCD Kondo effect \cite{Ozaki:2015sya}.} in a magnetic field through the feed down from the P-wave quarkonia.
In particular, by the HPBE, the wave function of a quarkonium with $L=1$ is separated into the ``polarized" components of $L_z=\pm1$ and $L_z=0$ in a sufficiently strong magnetic field, and such angular dependence of wave functions will lead to the azimuthal anisotropy of the quarkonium decays.

The purpose of this paper is to investigate the anisotropic decays of quarkonia from the HPBE and, in particular, to derive analytic formulas in the PB limit.
As representative examples, we focus on the radiative decays.
To this end, we apply the potential nonrelativistic QCD (pNRQCD) framework \cite{Pineda:1997bj,Brambilla:1999xf} (see Refs.~\cite{Brambilla:2004jw,Pineda:2011dg} for a review), which has been successfully applied to radiative decays of bottomonia \cite{Brambilla:2005zw,Brambilla:2012be,Pineda:2013lta}.
Such an approach provides us with systematic understanding based on the power counting scheme, and the extension to the higher order corrections will be straightforward.
To study decay properties in the PB limit, unlike the usual manner, we propose a novel approach that implements polarized wave functions into the usual pNRQCD framework.

Notice that, as a similar problem, the decay properties from the $J_z$ polarization of the P-wave quarkonia {\it in vacuum} were studied in Refs.~\cite{Faccioli:2011be,Shao:2012fs,Shao:2014fca}.
Such a polarization is distinguished only by $J_z$: $\chi_{c(b)1} (J_z=0,\pm1)$ and $\chi_{c(b)2} (J_z=0,\pm1,\pm2)$.
This is essentially different from polarization by the HPBE in finite magnetic fields, where each eigenstate has a specific configuration with $L_z$ and $S_z$.

This paper is organized as follows.
In Sec.~\ref{Sec_rev}, we briefly review the HPBE and summarize the expected wave function bases in the strong-field limit.
In Sec.~\ref{Sec_pNRQCD}, we formalize the pNRQCD with polarized wave functions.
In Sec.~\ref{Sec_decay}, the scattering amplitudes and decay widths are calculated and anisotropic E1 and M1 decays from P-wave quarkonia are discussed.
Section~\ref{Sec_conc} is devoted to our conclusion and outlook.

\section{Brief review of the HPBE} \label{Sec_rev}
Here, we briefly review the HPBE found in Ref.~\cite{Iwasaki:2018pby}.
In vacuum, the P-wave quarkonia are classified by spin-singlet $h_{c(b)}$ ($^1 \! P_1$) and spin-triplets $\chi_{c(b)0}$ ($^3 \! P_0$), $\chi_{c(b)1}$ ($^3 \! P_1$), and  $\chi_{c(b)2}$ ($^3 \! P_2$), where the total angular momentum $J=L+S$, orbital angular momentum $L$, and spin angular momentum $S$ for a $^{2S+1} \! P_J$ state are the good quantum numbers due to the spherical symmetry of the vacuum (note that $L_z$ and $S_z$ are not conserved owing to the LS and tensor coupling).

When a magnetic field along the $z$ direction is applied, the spherical symmetry is broken and only $J_z$ is strictly conserved.
When the magnetic field is stronger than the spin-orbit splitting, i.e., the PB region, 
$L_z$ and $S_z$ are also conserved approximately.\footnote{Precisely speaking, the existence of the tensor coupling mixes $L_z$ and $S_z$ even in the PB limit.}
The eigenstates can be represented by the {\it PB configuration} as follows:\footnote{In Ref.~\cite{Iwasaki:2018pby}, an alternative basis, $\Psi_{L_z;S_{1z}S_{2z}}(\rho,z,\phi) = \Phi_{L_z}(\rho, z) Y_{1 L_z}(\theta,\phi) \chi(S_{1z}, S_{2z})$, was introduced, where $S_{1z}$ and $S_{2z}$ are the $z$ components of the spin of the quark and antiquark.
} 
\e{
\Psi_{L_z;SS_z}(\rho,z,\phi) = \Phi_{L_z}(\rho, z) Y_{L L_z}(\theta,\phi) \chi_{S S_z},
\label{eq.PBconf}
}
where $\tan\theta=\rho/z$.
$\Phi_{L_z}$, $Y_{L L_z}$, and $\chi_{S S_z}$ in the right-hand side are the spatial, angular, and spin components of the wave function, respectively.

By using the representation of $Y_{L L_z} \chi_{S S_z}$ and the Clebsch-Gordon coefficients, the wave functions of the P-wave quarkonia in vacuum are summarized in Table \ref{Tab_CG_vac}.\footnote{In vacuum,
each eigenstate of (unpolarized) P-wave quarkonia is represented as the superposition of the states with each $J_z$, so that its wave function has the spherical symmetry.}
In the spin-triplet states ($^3 \! P_0$, $^3 \! P_1$, and $^3 \! P_2$) with each $J_z$, $Y_{L L_z} \chi_{S S_z}$ bases are mixed with each other by the LS and/or tensor coupling.
On the other hand, the spin-singlet states ($^1 \! P_1$) with each $J_z$ have no mixing.

The wave functions in the PB limit are shown in Table \ref{Tab_CG_PB}, which was numerically confirmed by the constituent quark model in a magnetic field \cite{Iwasaki:2018pby}.
Although these bases are approximate forms justified only in the strong magnetic fields, these will be useful for qualitative investigation by simplified wave functions.
The properties of the P-wave quarkonia in the PB limit are as follows:
\begin{itemize}
\item[(a)] The states with $S_z=0$, $Y_{10} \chi_{00}$ and $Y_{10} \chi_{10}$ for $J_z=0$ and $Y_{1\pm1}  \chi_{00}$ and $Y_{1\pm 1} \chi_{10}$ for $J_z=\pm1$, are mixed by the coupling between quark magnetic moments and magnetic field.\footnote{The coupling term between a magnetic field $\bm{B}=(0,0,B)$ and the quark magnetic moments $\bm{\mu}_i$ is given by $ \sum_i -\bm{\mu}_i \cdot \bm{B}$, where $\bm{\mu}_i = g q_i \bm{S}_i/2m_i$ with the Land$\acute{\mathrm{e}}$ $g$ factor, the electric charge $q_i$, the spin operator $\bm{S}_i$, and the mass $m_i$ for the quarks, 
Through this term, the off-diagonal matrix element between the spin eigenstates, $\chi_{00}$ and $\chi_{10}$, for a meson [or, equivalently, $\frac{1}{\sqrt{2}}(\ket{ \uparrow \downarrow} - \ket{\downarrow \uparrow})$ and $\frac{1}{\sqrt{2}}(\ket{ \uparrow \downarrow} + \ket{\downarrow \uparrow})$ for the spin eigenstates of the quark-antiquark pair] becomes nonzero \cite{Machado:2013rta,Alford:2013jva}:
\begin{equation}
\bra{Y_{LL_z} \chi_{10}} -(\bm{\mu}_1 + \bm{\mu}_2) \cdot \bm{B} \ket{Y_{LL_z} \chi_{00}} = -\frac{gB}{4} \left( \frac{q_1}{m_1} - \frac{q_2}{m_2} \right),
\end{equation}
since $q_1/m_1 \neq q_2/m_2$ for all of the (neutral and charged) mesons.
Note that the coupling term with the orbital angular momentum, given by $\bm{L} \cdot \bm{B}$, cancels for quarkonia (see, e.g., Eq.~(20) in Ref.~\cite{Alford:2013jva}).
}
\item[(b)] $Y_{11} \chi_{1-1}$ and $Y_{1-1} \chi_{11}$ for $J_z=0$ are mixed by the tensor coupling even in the PB limit.
\end{itemize}
Thus, to investigate these eigenstates, we need to construct an effective field theory with hadronic degrees of freedom (d.o.f.) in the PB limit.

\begin{table}[t!]
\centering
\caption{Wave-function bases of P-wave quarkonia {\it in vacuum}.}
\begin{tabular}{lcc}
\hline\hline
States & $J_z$ & Basis $(Y_{LL_z} \chi_{S S_z})$ \\
\hline
$h_{c(b)} \ (^1 \! P_1)$ & $0$    & $ Y_{10}  \chi_{00} $ \\
        & $\pm1$ & $ Y_{1\pm1} \chi_{00} $ \\
$\chi_{c(b)0} \ (^3 \! P_0)$ & $0$    & $ \frac{1}{\sqrt{3}} [Y_{11} \chi_{1-1} - Y_{10} \chi_{10} +Y_{1-1} \chi_{11}] $ \\
$\chi_{c(b)1} \ (^3 \! P_1)$ & $0$    & $ \frac{1}{\sqrt{2}} [Y_{1-1} \chi_{11} - Y_{11} \chi_{1-1}]$ \\
        & $\pm1$ & $ \pm \frac{1}{\sqrt{2}} [Y_{10} \chi_{1\pm1} - Y_{1\pm 1} \chi_{10}]$ \\
$\chi_{c(b)2} \ (^3 \! P_2)$ & $0$ & $ \frac{1}{\sqrt{6}} [Y_{11} \chi_{1-1} + 2 Y_{10} \chi_{10} + Y_{1-1} \chi_{11} ] $ \\
  & $\pm1$ & $ \frac{1}{\sqrt{2}} [Y_{1 \pm1} \chi_{10} + Y_{10} \chi_{1\pm1}] $ \\
  & $\pm2$ & $ Y_{1 \pm1} \chi_{1\pm1} $ \\
\hline\hline
\end{tabular}
\label{Tab_CG_vac}
\end{table}

\begin{table}[t!]
\centering
\caption{Wave-function bases of P-wave quarkonia {\it in the PB limit} \cite{Iwasaki:2018pby}.}
\begin{tabular}{cc}
\hline\hline
$J_z$ & Basis $(Y_{LL_z} \chi_{S S_z})$ \\
\hline
$0$  & $ \frac{1}{\sqrt{2}} [Y_{10}  \chi_{00} + Y_{10} \chi_{10}] $ \\
$0$  & $ \frac{1}{\sqrt{2}} [Y_{11} \chi_{1-1} + Y_{1-1} \chi_{11}] $ \\
$0$  & $ \frac{1}{\sqrt{2}} [Y_{11} \chi_{1-1} - Y_{1-1} \chi_{11}] $ \\
$0$  & $ \frac{1}{\sqrt{2}} [Y_{10}  \chi_{00} - Y_{10} \chi_{10}] $ \\
\hline
$\pm1$ & $ \frac{1}{\sqrt{2}} [Y_{1\pm1}  \chi_{00} +  Y_{1\pm 1} \chi_{10}] $ \\
$\pm1$ & $ Y_{10} \chi_{1\pm1}$  \\
$\pm1$ & $ \frac{1}{\sqrt{2}} [Y_{1\pm1}  \chi_{00} -  Y_{1\pm 1} \chi_{10}] $\\
\hline
$\pm2$ & $ Y_{1 \pm1} \chi_{1\pm1} $ \\
\hline\hline
\end{tabular}
\label{Tab_CG_PB}
\end{table}

\section{pNRQCD in the PB limit} \label{Sec_pNRQCD}
In this section, we briefly review pNRQCD and construct a novel formalism for quarkonia in the PB limit.

\subsection{pNRQCD Lagrangian}
In pNRQCD \cite{Pineda:1997bj,Brambilla:1999xf}, an effective field theory is constructed based on the hierarchy of energy scales for heavy-quark bound states: $m>mv>mv^2$, where $m$ and $v \, (\sim 1/m)$ are the heavy-quark mass and velocity, respectively.
Such a scale separation is reasonable for bottomonia ($v^2 \sim 0.1$) and might also be applied to charmonia ($v^2 \sim 0.3$).
The relevant d.o.f. are heavy quarks in the potential region ($p^0 \sim mv^2, |{\bf p}| \sim mv$) and gluons in the ultrasoft region ($p^0, |{\bf p}| \sim mv^2$), while the harder d.o.f. in the hard ($p^0, |{\bf p}| \sim m$) and soft ($p^0, |{\bf p}| \sim mv$) region are integrated out.
The Lagrangian contains the color-singlet field $\mathrm{S} ({\bf R},{\bf r})$ and the color-octet field $\mathrm{O} ({\bf R},{\bf r})$ with the center-of-mass coordinate ${\bf R}$ and the relative distance ${\bf r}$ between quark and antiquark, which is given by
\begin{eqnarray}
\mathcal{L}_{\mathrm{pNRQCD}} &=&  \int d^3 r \mathrm{Tr} \left\{ \mathrm{S}^\dag \left( i\partial_0 +\frac{\mbox{\boldmath $\nabla$}^2}{4m} + \frac{\mbox{\boldmath $\nabla$}_r^2}{m} + \cdots - V_S \right) \mathrm{S} \right\} \nonumber \\
&& + \mathcal{L}_\mathrm{octet} + \mathcal{L}_{\gamma \mathrm{pNRQCD}},
\end{eqnarray}
where $\mbox{\boldmath $\nabla$}^i = \partial/\partial {\bf R}^i$, $\mbox{\boldmath $\nabla$}_r^i = \partial/\partial {\bf r}^i$, and 
$V_S$ is the potential term for the color singlet.
For weakly coupled quarkonia, at the leading order of $\alpha_s$, it is just the Coulomb potential ($V_S^{(0)}=-C_F\frac{\alpha_s}{r}$), where $C_F=(N_c^2-1)/2N_c=4/3$ and $N_c=3$.
For strongly coupled quarkonia, it is determined by the matching to NRQCD.
The trace is over the color and spin indices.
$\mathcal{L}_\mathrm{octet}$ contains the kinetic and potential terms for the color-octet states and the mixing terms between singlet and octet states.
$\mathcal{L}_{\gamma \mathrm{pNRQCD}}$ is the interaction term of quarkonia and photons, where the photon energy is limited to be smaller than the scale of $mv^2$: $k_\gamma \simle mv^2$ (photons with larger energy are integrated out).
The final observables, such as decay widths, are represented as the $v^2$ expansion, and we will neglect the corrections with the higher-order $v^2$ term.

The quarkonium eigenstate of the full pNRQCD Hamiltonian, $ \ket{H ( {\bf P}, \lambda)}$ with the center-of-mass momentum ${\bf P}$ and the polarization $\lambda$, is normalized by
\begin{equation}
\langle H ({\bf P}^\prime, \lambda^\prime) \ket{H ( {\bf P}, \lambda)} = \delta_{\lambda \lambda^\prime} (2\pi)^3  \delta^3({\bf P}- {\bf P}^\prime).
\end{equation}
The eigenstates at the leading order pNRQCD are defined by
\begin{eqnarray}
\ket{H ( {\bf P}, \lambda)}^{(0)} &=&  \int d^3 R \int d^3 r e^{i {\bf P} \cdot {\bf R}} \mathrm{Tr} \{ \phi_{H(\lambda)}^{(0)} ({\bf r}) \mathrm{S}^\dag ({\bf r},{\bf R}) | \mathrm{vac} \rangle \}. \nonumber\\
\end{eqnarray}
The wave function $ \phi_{H(\lambda)}^{(0)} ({\bf r})$ is an eigenfunction of the leading-order pNRQCD Hamiltonian: $h^{(0)} \equiv - \mbox{\boldmath $\nabla$}_r^2/m + V_S^{(0)}$).

On the other hand, the one photon state $ \ket{\gamma ( {\bf k}, \sigma)}$ with momentum ${\bf k}$ and the polarization $\sigma$ is normalized by 
\begin{equation}
\langle \gamma ({\bf k}^\prime, \sigma^\prime) \ket{\gamma ( {\bf k}, \sigma)} = \delta_{\sigma \sigma^\prime} 2k (2\pi)^3 \delta^3({\bf k}- {\bf k}^\prime).
\end{equation}

\subsection{(Unpolarized) bases of wave functions in vacuum}

\subsubsection{S wave in vacuum}
First we define the wave functions for $L=0$ in vacuum [{\it e.g.}, $\eta_c$ ($^1S_0$) and $J/\psi$ ($^3S_1$)] \cite{Brambilla:2005zw}:
\begin{eqnarray}
&& \phi^{(0)}_{n^1S_0} ({\bf r}) = \sqrt{ \frac{1}{8\pi}} R_{n0}(r), \label{base_1S0} \\
&& \phi^{(0)}_{n^3S_1} ({\bf r}) = \sqrt{ \frac{1}{8\pi}} R_{n0}(r) {\boldsymbol \sigma} \cdot {\bf e}_{n^3S_1}(\lambda),  \label{base_3S1}
\end{eqnarray}
where ${\boldsymbol \sigma} = (\sigma^x, \sigma^y, \sigma^z)$ is the Pauli vector.
The polarization vector of the $n^3S_1$ states, $ {\bf e}_{n^3S_1}(\lambda)$, is normalized as $ {\bf e}_{n^3S_1}^\ast (\lambda) \cdot {\bf e}_{n^3S_1}(\lambda^\prime) = \delta_{\lambda \lambda^\prime}$.
The factor of $\sqrt{1/8\pi}$ is the normalization factor, and $R_{nL}(r)$ is the spatial part of the wave functions.
These wave functions will be used for the calculations of E1 decays from a P-wave into an S-wave in vacuum in Sec.~\ref{Sec_E1}.

\subsubsection{P wave in vacuum}
The wave functions for $L=1$ in vacuum [{\it e.g.}, $h_c$ ($^1P_1$), $\chi_{c0}$ ($^3P_0$), $\chi_{c1}$ ($^3P_1$), and $\chi_{c2}$ ($^3P_2$)] are defined as \cite{Brambilla:2005zw}
\begin{eqnarray}
&& \phi^{(0)}_{n^1P_1(\lambda)} ({\bf r}) = \sqrt{ \frac{3}{8\pi}} R_{n1}(r) {\bf e}_{n^1P_1}(\lambda) \cdot \hat{\bf r},  \label{base_1P1} \\
&& \phi^{(0)}_{n^3P_0} ({\bf r}) = \sqrt{ \frac{1}{8\pi}} R_{n1}(r) {\boldsymbol \sigma} \cdot \hat{\bf r},  \label{base_3P0} \\
&& \phi^{(0)}_{n^3P_1(\lambda)} ({\bf r}) = \sqrt{ \frac{3}{16\pi}} R_{n1}(r) {\boldsymbol \sigma} \cdot ( \hat{\bf r} \times {\bf e}_{n^3P_1}(\lambda) ),  \label{base_3P1} \\
&& \phi^{(0)}_{n^3P_2(\lambda)} ({\bf r}) = \sqrt{ \frac{3}{8\pi}} R_{n1}(r) {\boldsymbol \sigma}^i h^{ij}_{n^3P_2} \hat{\bf r}^j,  \label{base_3P2}
\end{eqnarray}
where $ \hat{\bf r}= {\bf r}/|{\bf r}|  = (\hat{x}, \hat{y}, \hat{z})$ is the three-component unit vector in position space.
The polarization vector and tensor are normalized as $ {\bf e}_{n^1P_1}^\ast (\lambda) \cdot {\bf e}_{n^1P_1}(\lambda^\prime) =  {\bf e}_{n^3P_1}^\ast (\lambda) \cdot {\bf e}_{n^3P_1}(\lambda^\prime) = \delta_{\lambda \lambda^\prime}$ and $ h_{n^3P_2}^{ij \ast} (\lambda) h_{n^3P_2}^{ij} (\lambda^\prime) = \delta_{\lambda \lambda^\prime}$, respectively.

\subsection{Polarized bases of wave functions}
Next, we define {\it polarized} wave functions motivated by PBE in relatively strong magnetic fields (PB region), where the good quantum numbers are $J_z$, $L_z$, and $S_z$ (if we neglect the LS and tensor coupling).

\subsubsection{P wave ($J_z=0$) in the PB limit}
We define the P-wave functions in the PB limit.
For $J_z=0$,
\begin{eqnarray}
&& \phi^{(0)}_{L_z=0;(S,S_z)=(0,0)} ({\bf r}) = \sqrt{\frac{3}{8\pi}} R_{n1}(r) \hat{z}, \label{pol_base1} \\ 
&& \phi^{(0)}_{L_z=0;(S,S_z)=(1,0)} ({\bf r}) = \sqrt{\frac{3}{8\pi}} R_{n1}(r) \hat{z} {\sigma}^{z}, \label{pol_base2} \\
&& \phi^{(0)}_{L_z=+1;(S,S_z)=(1,-1)} ({\bf r}) = \sqrt{\frac{3}{8\pi}} R_{n1}(r) \hat{r}^+ {\sigma}^{-}, \label{pol_base3} \\
&& \phi^{(0)}_{L_z=-1;(S,S_z)=(1,+1)} ({\bf r}) = \sqrt{\frac{3}{8\pi}} R_{n1}(r) \hat{r}^-{\sigma}^{+}, \label{pol_base4}
\end{eqnarray}
where ${\sigma}^{\pm} = (\sigma^x \pm i \sigma^y)/ \sqrt{2}, \hat{r}^{\pm} = (\hat{x} \pm i \hat{y})/ \sqrt{2}$.

\subsubsection{P wave ($J_z=\pm1$) in the PB limit}
For $J_z=\pm1$,
\begin{eqnarray}
&& \phi^{(0)}_{L_z=\pm 1;(S,S_z)=(0,0)} ({\bf r}) = \sqrt{\frac{3}{8\pi}} R_{n1}(r) \hat{r}^{\pm}, \label{pol_base5} \\
&& \phi^{(0)}_{L_z=\pm 1;(S,S_z)=(1,0)} ({\bf r}) = \sqrt{\frac{3}{8\pi}} R_{n1}(r) \hat{r}^{\pm} {\sigma}^{z}, \label{pol_base6} \\
&& \phi^{(0)}_{L_z=0;(S,S_z)=(1,\pm 1)} ({\bf r}) = \sqrt{\frac{3}{8\pi}} R_{n1}(r) \hat{z} {\sigma}^{\pm}. \label{pol_base7}
\end{eqnarray}

\subsubsection{P wave ($J_z=\pm2$) in the PB limit}
For $J_z=\pm2$,
\begin{equation}
\phi^{(0)}_{L_z=\pm 1;(S,S_z)=(1,\pm 1)} ({\bf r}) = \sqrt{\frac{3}{8\pi}} R_{n1}(r) \hat{r}^{\pm} {\sigma}^{\pm}. \label{pol_base8}
\end{equation}

\subsection{Redefinition of photon polarization vector}

\begin{figure}[t!]
    \centering
    \includegraphics[clip, width=1.0\columnwidth]{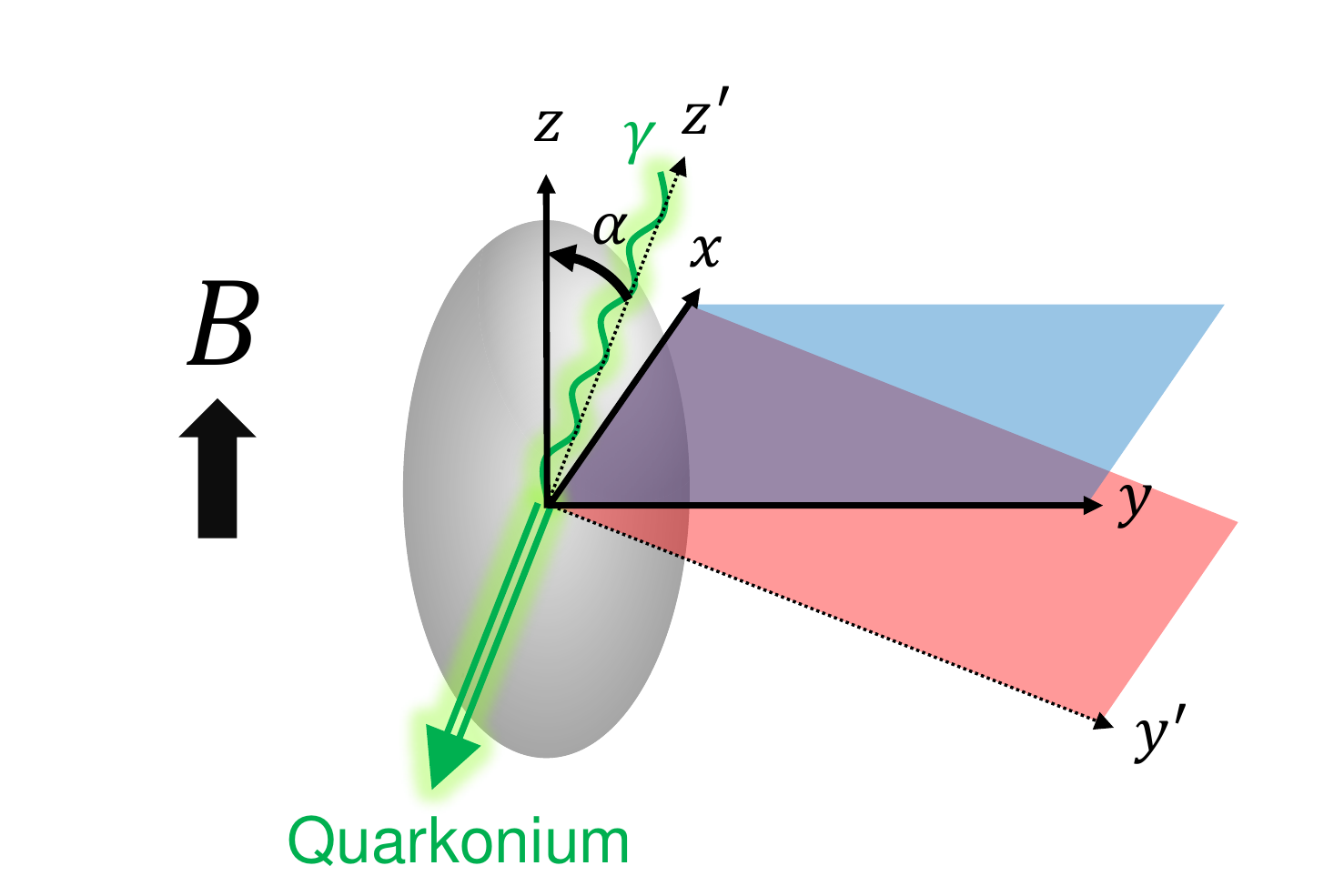}
    \caption{Definition of decay angles $\alpha$.
The initial state of quarkonium at rest decays into a photon and the final state of quarkonium.}
    \label{fig_decay}
\end{figure}

Now our quantization axis $z$ is parallel to the magnetic field.
Since, in general, the directions of the magnetic field and photon emission are different from each other, the quantization axis for the polarization vector of the photon is required to be redefined. 
The original polarization vector with the quantization axis $z^\prime$ parallel to the photon momentum is defined as ${\boldsymbol \epsilon}_\mathrm{org} \, (\sigma=\pm) = \frac{1}{\sqrt{2}} (\pm 1, -i, 0)$.
This is rotated by an angle $\alpha$ around the original $x$ axis, as shown in Fig.~\ref{fig_decay}:
\begin{eqnarray}
 {\boldsymbol \epsilon} \, (\sigma=\pm) &=& \frac{1}{\sqrt{2}} (\pm 1, -i, 0) \left(
    \begin{array}{ccc}
      1 & 0 & 0 \\
      0 & \cos \alpha & \sin \alpha \\
      0 & -\sin \alpha & \cos \alpha
    \end{array}
  \right) \nonumber \\
&=& \frac{1}{\sqrt{2}} (\pm 1, -i \cos \alpha,  -i \sin \alpha). \label{photon_pol}
\end{eqnarray}
Note that when $\alpha=0$, this vector agrees with the original one, which corresponds to the photon emission parallel to the $z$ axis (and the magnetic field).
Also, the case for $\alpha=\pi/2$ corresponds to the emission on the $x$-$y$ plane.

\section{Radiative decays in the PB limit} \label{Sec_decay}
In this section, we investigate E1 and M1 decays from the quarkonium eigenstates in PB limit.

\subsection{E1 decays in the PB limit} \label{Sec_E1}
E1 transitions change the orbital angular momentum of quarkonia by $\Delta L = \pm 1$, while they conserve the spin angular momentum: $\Delta S=0$.
In vacuum, the E1 decay processes such as $h_c \to \eta_c \gamma$ and $\chi_c \to J/\psi \gamma$ are possible.
On the other hand, in a strong magnetic field, the wave functions of the P-wave quarkonia are modified by the HPBE, and we have many decay channels.

The E1 transition operator at leading order is given by \cite{Brambilla:2012be}
\begin{equation}
\mathcal{L}_\mathrm{E1}^{(0)} =  ee_Q \int d^3 r \mathrm{Tr} \{\mathrm{S}^\dag {\bf r} \cdot {\bf E}^\mathrm{em} \mathrm{S} \},
\end{equation}
where $ee_Q$ is the quark electric charge ($e$ is the elementary charge, $e_c=2/3$, and $e_b=-1/3$) and ${\bf E}^\mathrm{em}$ is the external electric field.
Using this Lagrangian, the scattering amplitude from an initial quarkonium state $| H ( {\bf 0}, \lambda ) \rangle$ with $(M_H, \bf{P} = \bf{0})$ at rest into a final state $| H^\prime ( {\bf P}^\prime, \lambda^\prime ) \rangle$ with $(M_{H^\prime}, {\bf P}^\prime)$ and one photon state $|\gamma ( {\bf k}, \sigma ) \rangle$ with energy $k_\gamma = |{\bf k}| = (M_H^2-M_{H^\prime}^2)/2M_H$ is
\begin{eqnarray}
&& \mathcal{A}_\mathrm{E1}^{(0)} (2\pi)^3 \delta^3 ({\bf P}^\prime + {\bf k}) \nonumber\\
&& = - \langle H^\prime ( {\bf P}^\prime, \lambda^\prime ) \gamma ( {\bf k}, \sigma) | \int d^3R \, \mathcal{L}_{\mathrm{E1}}^{(0)} | H ( {\bf 0}, \lambda ) \rangle.
\end{eqnarray}
To calculate the production amplitude of the photon state, we use the external electric field projection on the one photon state:
\begin{equation}
\langle \gamma({\bf k}, \sigma) | {\bf E}^\mathrm{em} | \mathrm{vac} \rangle = -ik {\boldsymbol \epsilon}^{\ast} (\sigma) e^{-i {\bf k} \cdot {\bf R}},
\end{equation}
where we note that the photon polarization vector ${\boldsymbol \epsilon} (\sigma) $ is redefined in Eq.~(\ref{photon_pol}).

\subsubsection{Decays from only the HPBE deformation}
First, for simplicity, we consider the decays from the wave functions taking into account only the deformation by the HPBE.
In this case, we neglect the deformation from other sources, such as quark Landau levels.

For the decays from the $L_z=0$ basis [(\ref{pol_base1}), (\ref{pol_base2}), and (\ref{pol_base7})] into the S-wave basis [(\ref{base_1S0}) and (\ref{base_3S1})], the amplitude squared, which is summed over the photon polarization $\sigma$, is
\begin{eqnarray}
&& \sum_\sigma  |\mathcal{A}_{\ket{L_z=0} \to n^{\prime }S  \gamma} |^2 = \nonumber\\
&& \frac{4\pi \alpha_{em} e_Q^2 k^2}{3} \left[ \int_0^\infty dr r^3 R_{n^\prime L^\prime} (r) R_{nL} (r) \right]^2 \sin^2 \alpha,
\end{eqnarray}
where $\alpha_{em} = e^2/4\pi$ is the fine structure constant.
For the $L_z=\pm1$ basis [(\ref{pol_base3})--(\ref{pol_base6}), and (\ref{pol_base8})],
\begin{eqnarray}
&& \sum_\sigma  |\mathcal{A}_{\ket{L_z=\pm1} \to n^{\prime }S  \gamma} |^2 = \nonumber\\
&& \frac{2\pi \alpha_{em} e_Q^2 k^2}{3} \left[ \int_0^\infty dr r^3 R_{n^\prime L^\prime} (r) R_{nL} (r) \right]^2 ( \cos^2 \alpha + 1). \nonumber\\
\end{eqnarray}

Finally, we obtain the analytic formulas for the E1 decay widths:
\begin{eqnarray}
\Gamma_{\ket{L_z=0} \to n^{\prime 3}S_1  \gamma} &=& \frac{1}{8\pi^2} \left( 1-\frac{k_\gamma}{M_H} \right) \int_0^\infty dk k \nonumber\\
&& \times \int d \Omega(\hat{{\bf k}}) \delta(k-k_\gamma) \sum_\sigma |\mathcal{A}_{\ket{L_z=0} \to n^{\prime }S  \gamma} |^2  \nonumber\\
&=& \frac{2}{3} \alpha_{em} e_Q^2 k_\gamma^3 \left[ \int_0^\infty dr r^3 R_{n^\prime L^\prime} (r) R_{nL} (r) \right]^2 \nonumber\\
&& \left(1- \frac{k_\gamma}{M_H} \right)  \sin^2 \alpha, \label{PB_E1_width1}
\end{eqnarray}
\begin{eqnarray}
\Gamma_{\ket{L_z=\pm1} \to n^{\prime 3}S_1  \gamma} &=& \frac{1}{8\pi^2} \left( 1-\frac{k_\gamma}{M_H} \right) \int_0^\infty dk k \nonumber\\
&& \times \int d \Omega(\hat{{\bf k}}) \delta(k-k_\gamma) \sum_\sigma |\mathcal{A}_{\ket{L_z=\pm1} \to n^{\prime} S  \gamma} |^2  \nonumber\\
&=& \frac{1}{3} \alpha_{em} e_Q^2 k_\gamma^3 \left[ \int_0^\infty dr r^3 R_{n^\prime L^\prime} (r) R_{nL} (r) \right]^2 \nonumber\\
&& \times \left(1- \frac{k_\gamma}{M_H} \right) ( \cos^2 \alpha + 1). \label{PB_E1_width2}
\end{eqnarray}

Note that the widths of the E1 decays from the eigenstates in vacuum ($^1 \! P_1 \to ^1 \!\! S_0$ and $^3 \! P_0, ^3 \! P_1, ^3 \! P_2 \to ^3 \! \! S_1$) have a factor of $4/9$, which have no angler dependence, as shown in Appendix \ref{App_vac}.

From these formulas, our findings are as follows:
\begin{itemize}
\item[(a)] The factors of $ \sin^2 \alpha$ and $ \cos^2 \alpha + 1$ indicate the angular dependence of the decay width.
\item[(b)] The E1 decay width by the {\it $L_z$-conserved} transition, from the basis $\ket{L_z=0}$ for $\alpha=\pi/2$, has the factor of $2/3$, which is larger than that in vacuum, so that the decays perpendicular to the magnetic field are enhanced, while the decay of $\ket{L_z=0}$ for $\alpha=0, \pi$ is forbidden.
\item[(c)] The E1 decay widths by the {\it $L_z$-changed} transitions, from the bases $\ket{L_z=\pm 1}$ for $\alpha=0, \pi$ has the factor of $2/3$, which is larger than that in vacuum, so that the decays parallel to the magnetic field is enhanced.
The decay width of $\ket{L_z=\pm1}$ for $\alpha=\pi/2$ has the factor of $1/3$, and the decays are suppressed.
\item[(d)] The spin of the final spin-triplet S-wave quarkonium states is automatically polarized although the original S-wave basis (\ref{base_3S1}) is not polarized.
\end{itemize}

It should be noted that these formulas reflect only the deformation of P-wave wave functions by the HPBE.
These can be justified when the magnetic field is weaker than a relevant scale for deformation of the S wave and P wave by other magnetic effects.
In fact, the deformation in the ground state of the S-wave quarkonia is induced by Landau levels of heavy quarks, and it is relevant for the heavy-quark mass scale \cite{Suzuki:2016kcs,Yoshida:2016xgm}, which indicates that these formulas can be safely applied for $\sqrt{|eB|} \ll m$.
For the P-wave and excited states, the deformation by Landau levels is expected to be stronger.
However, we emphasize that the relevant scale for the HPBE is the LS coupling, and the HPBE can occur even in a much weaker magnetic field than relevant fields for other effects.
Therefore, the formulas, Eqs.~(\ref{PB_E1_width1}) and (\ref{PB_E1_width2}), are justified in the region of relatively weak fields, $V_{LS} \ll \sqrt{|eB|} \ll V_{other}$.

\subsubsection{Decays from general deformation}
Next, we consider decays taking into account a more general deformation of the wave functions, which includes quark Landau levels and the deformation of the potential between the quarks ({\it e.g.}, see Ref.~\cite{Bonati:2014ksa}) as well as the HPBE.
In this case, $\theta$ in the spherical coordinates $(r,\theta,\phi)$ is not integrated out from scattering amplitudes, and then they are represented by $\rho$ and $z$ in the cylindrical coordinates $(\rho,z,\phi)$.
Therefore, the spatial parts of the wave functions should be represented not by $R_{nL}(r)$ but rather by $R_{nL}^\prime(\rho,z)$.

\begingroup
\renewcommand{\arraystretch}{1.5}
\begin{table*}[tbh!]
\centering
\caption{Summary of E1 decays from P-wave quarkonia in the PB limit into S-wave quarkonia {\it in vacuum}.
$\alpha=0$ ($\alpha=\pi/2$) corresponds to the direction parallel (perpendicular) to the magnetic field.
$Y_{00} \chi_{00}$, $Y_{00} \chi_{10}$, $Y_{00} \chi_{11}$, and $Y_{00} \chi_{1-1}$ correspond to the $^1 \! S_0$, $ ^3 \! S_1 (S_z=0)$, $ ^3 \! S_1 (S_z=+1)$, and $ ^3 \! S_1 (S_z=-1)$ states, respectively.}
\begin{tabular}{ccccc}
\hline\hline
$\Delta J_z$ & Decay process & Decay factor & $\alpha=0,\pi$ & $\alpha=\pi/2
$ \\
\hline
$0 \to 0$  & $ \frac{1}{\sqrt{2}} [Y_{10}  \chi_{00} \pm Y_{10} \chi_{10}] \to Y_{00} \chi_{00},Y_{00} \chi_{10}$ & \large $\frac34 \sin^2 \alpha$ & $0$ & \large $  \frac3{4}$  \\
$0 \to -1$  & $ \frac{1}{\sqrt{2}} [Y_{11} \chi_{1-1} \pm Y_{1-1} \chi_{11}] \to Y_{00} \chi_{1-1}$ & \large $\frac{3}{32}( \cos^2 \alpha +1 )$ &\large $\frac{3}{16}$ & \large $\frac3{32}$ \\
$0 \to +1$  & $ \frac{1}{\sqrt{2}} [Y_{11} \chi_{1-1} \pm Y_{1-1} \chi_{11}] \to Y_{00} \chi_{11}$ & \large $\frac{3}{32}( \cos^2 \alpha +1) $ & \large $\frac{3}{16}$ & \large $\frac3{32}$ \\
\hline
$+1 \to 0$ & $ \frac{1}{\sqrt{2}} [Y_{11}  \chi_{00} \pm  Y_{11} \chi_{10}] \to Y_{00} \chi_{00}, Y_{00} \chi_{10}$ & \large $\frac{3}{32}( \cos^2 \alpha +1)$ & \large $\frac3{16}$ & \large $\frac3{32}$ \\
$\pm1 \to \pm1$ & $ Y_{10} \chi_{1\pm1} \to Y_{00} \chi_{1\pm1} $ & \large $\frac32 \sin^2 \alpha$ & $0$ & \large $\frac{3}{2}$  \\
$-1 \to 0$ & $ \frac{1}{\sqrt{2}} [Y_{1-1}  \chi_{00} \pm  Y_{1-1} \chi_{10}] \to Y_{00} \chi_{00}, Y_{00} \chi_{10} $ & \large $\frac{3}{32}( \cos^2 \alpha +1)$ &  \large $\frac3{16}$ & \large $\frac3{32}$ \\
\hline
$+2 \to +1$ & $ Y_{11} \chi_{11} \to Y_{00} \chi_{11} $ & \large $\frac{3}{16}(  \cos^2 \alpha +1 )$ & \large $\frac38$ & \large $\frac3{16}$ \\
$-2 \to -1$ & $ Y_{1-1} \chi_{1-1} \to Y_{00} \chi_{1-1} $ & \large $\frac{3}{16}( \cos^2 \alpha +1)$ & \large $\frac38$ & \large $\frac3{16}$ \\
\hline\hline
\end{tabular}
\label{Tab_decay_in_PB_E1}
\end{table*}
\endgroup

For the decays from the $L_z=0$ states [(\ref{pol_base1}), (\ref{pol_base2}), and (\ref{pol_base7})] into the S-wave states [(\ref{base_1S0}) and (\ref{base_3S1})], where $R_{nL}(r)$ is replaced by $R_{nL}^\prime(\rho,z)$, the amplitude squared, which is summed over the photon polarization $\sigma$, is

\begin{widetext}
\begin{eqnarray}
&& \sum_\sigma  |\mathcal{A}_{\ket{L_z=0} \to n^{\prime }S  \gamma} |^2 =
3\pi \alpha_{em} e_Q^2 k^2 
\left[ 
	\int_0^\infty \rho d\rho \, \int_{-\infty}^\infty dz 
	R_{n^\prime 0}^\prime(\rho,z) R_{n1}^\prime(\rho,z) \frac{z^2}{\sqrt{\rho^2+z^2}} 
\right]^2 \sin^2 \alpha.
\end{eqnarray}

For the $L_z=\pm1$ states [(\ref{pol_base3})--(\ref{pol_base6}), and (\ref{pol_base8})],
\begin{eqnarray}
&& \sum_\sigma  |\mathcal{A}_{\ket{L_z=\pm1} \to n^{\prime }S  \gamma} |^2 =  
\frac{3\pi \alpha_{em} e_Q^2 k^2}{8} 
\left[ 
	\int_0^\infty \rho^3 d\rho \, \int_{-\infty}^\infty d z
	R_{n^\prime 0}^\prime(\rho,z) R_{n1}^\prime(\rho,z) \frac{1}{\sqrt{\rho^2+z^2}}
\right]^2 ( \cos^2 \alpha + 1) .
\end{eqnarray}

Finally, we obtain the analytic formulas for the E1 decay widths:
\begin{eqnarray}
\Gamma_{\ket{L_z=0} \to n^{\prime 3}S_1  \gamma} &=& 
\frac32 \alpha_{em} e_Q^2 k_\gamma^3 
\left[ 
	\int_0^\infty \rho d\rho \, \int_{-\infty}^\infty dz 
	R_{n^\prime 0}^\prime(\rho,z) R_{n1}^\prime(\rho,z)  \frac{z^2}{\sqrt{\rho^2+z^2}} 
\right]^2 \left(1- \frac{k_\gamma}{M_H} \right)  \sin^2 \alpha, \label{PB_E1_width3}
\\
\Gamma_{\ket{L_z=\pm1} \to n^{\prime 3}S_1  \gamma} &=&
\frac3{16} \alpha_{em} e_Q^2 k_\gamma^3 
\left[ 
	\int_0^\infty \rho^3 d\rho \, \int_{-\infty}^\infty d z
	R_{n^\prime 0}^\prime(\rho,z) R_{n1}^\prime(\rho,z) \frac{1}{\sqrt{\rho^2+z^2}}
\right]^2 \left(1- \frac{k_\gamma}{M_H} \right) ( \cos^2 \alpha + 1). \label{PB_E1_width4}
\end{eqnarray}
\end{widetext}

Note that these formulas are more general forms than Eqs.~(\ref{PB_E1_width1}) and (\ref{PB_E1_width2}).
If we neglect deformation effects except for the HPBE, then we use the spherical coordinate, and $\theta$ can be integrated out.
Thus the formulas (\ref{PB_E1_width3}) and (\ref{PB_E1_width4}) become equal to Eqs.~(\ref{PB_E1_width1}) and (\ref{PB_E1_width2}) represented only by $r$, respectively.

\begingroup
\renewcommand{\arraystretch}{1.5}
\begin{table*}[tbh!]
\centering
\caption{Summary of E1 decays from P-wave quarkonia in the PB limit into S-wave quarkonia {\it in strong field limit}.
}
\begin{tabular}{cccccc}
\hline\hline
$\Delta J_z$ & Decay process & Decay factor & $\alpha=0,\pi$ & $\alpha=\pi/2$ \\
\hline
$0 \to 0$  & $ \frac{1}{\sqrt{2}} [Y_{10}  \chi_{00} \pm Y_{10} \chi_{10}] \to \frac{1}{\sqrt{2}} [Y_{00} \chi_{00} \pm Y_{00} \chi_{10}]$ & \large $\frac34 \sin^2 \alpha$ & $0$ & \large $  \frac3{4}$  \\
$0 \to -1$  & $ \frac{1}{\sqrt{2}} [Y_{11} \chi_{1-1} \pm Y_{1-1} \chi_{11}] \to Y_{00} \chi_{1-1}$ & \large $\frac{3}{32}(  \cos^2 \alpha +1)$ &\large $\frac{3}{16}$ & \large $\frac3{32}$ \\
$0 \to +1$  & $ \frac{1}{\sqrt{2}} [Y_{11} \chi_{1-1} \pm Y_{1-1} \chi_{11}] \to Y_{00} \chi_{11}$ & \large $\frac{3}{32} (  \cos^2 \alpha +1)$ &\large $\frac{3}{16}$ & \large $\frac3{32}$ \\
\hline
$+1 \to 0$ & $ \frac{1}{\sqrt{2}} [Y_{11}  \chi_{00} \pm  Y_{11} \chi_{10}] \to \frac{1}{\sqrt{2}}[Y_{00} \chi_{00} \pm Y_{00} \chi_{10}]$ & \large $\frac{3}{32}( \cos^2 \alpha +1)$ & \large $\frac3{16}$ & \large $\frac3{32}$ \\
$\pm1 \to \pm1$ & $ Y_{10} \chi_{1\pm1} \to Y_{00} \chi_{1\pm1} $ & \large $\frac32 \sin^2 \alpha$ & $0$ & \large $\frac{3}{2}$  \\
$-1 \to 0$ & $ \frac{1}{\sqrt{2}} [Y_{1-1}  \chi_{00} \pm  Y_{1-1} \chi_{10}] \to \frac{1}{\sqrt{2}}[Y_{00} \chi_{00} \pm Y_{00} \chi_{10}]$ & \large $\frac{3}{32}( \cos^2 \alpha +1)$ & \large $\frac3{16}$ & \large $\frac3{32}$ \\
\hline
$+2 \to +1$ & $ Y_{11} \chi_{11} \to Y_{00} \chi_{11} $ & \large $\frac{3}{16}(  \cos^2 \alpha +1 )$ & \large $\frac38$ & \large $\frac3{16}$ \\
$-2 \to -1$ & $ Y_{1-1} \chi_{1-1} \to Y_{00} \chi_{1-1} $ & \large $\frac{3}{16}(  \cos^2 \alpha +1 )$ & \large $\frac38$ & \large $\frac3{16}$ \\
\hline\hline
\end{tabular}
\label{Tab_decay_in_PB2_E1}
\end{table*}
\endgroup

\subsubsection{Decays from quarkonium eigenstates}

By using the formulas, Eqs.~(\ref{PB_E1_width3}) and (\ref{PB_E1_width4}), and the eigenstates defined in Table \ref{Tab_CG_PB}, we can discuss the decay properties from P-wave quarkonium eigenstates in the PB limit.
First, we see the E1 decay properties into the S-wave quarkonia {\it in vacuum}, which is summarized in Table \ref{Tab_decay_in_PB_E1}.
Here, we assumed that the spin wave functions of the S-wave quarkonia are the same as those in vacuum.
The justification of the assumption depends on the magnetic field dependences of the mixing rates between the spin eigenstates.

In the E1 decays, the spin angular momentum $S$ and its third component $S_z$ of the bases are conserved ($\Delta S = \Delta S_z = 0$), while the orbital angular momentum $L$ changes, $\Delta L \neq 0$.
For example, for the P-wave charmonia with $J_z=0$, we have the four eigenstates in the series with the same principal quantum number $n$ (see Ref.~\cite{Iwasaki:2018pby} for the mass spectra).
The first (lightest) and fourth (heaviest) states, $\frac{1}{\sqrt{2}} [Y_{10}  \chi_{00} \pm Y_{10} \chi_{10}]$, go to both $\eta_c$ and $J/\psi (S_z=0)$ by the ratio of $1:1$, and the decay factor is $\frac{3}{4} \sin^2 \alpha$.
Therefore, their production parallel to the magnetic field is forbidden, while that on the transverse plane is enhanced.
On the other hand, the second and third ones ($\frac{1}{\sqrt{2}} [Y_{11} \chi_{1-1} \pm Y_{1-1} \chi_{11}]$) go to both $J/\psi (S_z=+1)$ and $J/\psi (S_z=-1)$ by the ratio of $1:1$, and the angle factor is $\frac{3}{32} (\cos^2 \alpha + 1)$.

For $J_z=+1$, we have the three eigenstates.
The first (lightest) and third (heaviest) states, $\frac{1}{\sqrt{2}} [Y_{11}  \chi_{00} \pm  Y_{11} \chi_{10}]$, go to $\eta_c$ and $J/\psi (S_z=0)$, and the decay factor is $\frac{3}{32} ( \cos^2 \alpha + 1)$, so that their production parallel to the magnetic field is enhanced, and that on the transverse plane is relatively suppressed.
On the other hand, the second state $Y_{10} \chi_{11}$ goes to only $J/\psi (S_z=+1)$, and the decay factor is $\frac{3}{2} \sin^2 \alpha$, where the factor of $\frac{3}{2}$, which is two times more than the $0 \to 0$ channel, comes from ``pure" (nonmixed) bases of the initial quarkonia.
For $J_z=+2$, we have the only one eigenstate, and it goes to only $J/\psi (S_z=+1)$.
For a negative $J_z$, the discussion is the same as that for a positive $J_z$.

Next, we comment on how to take into account the mixing between spin eigenstates for the S wave.
In Table \ref{Tab_decay_in_PB_E1}, we have used the S-wave eigenstates {\it in vacuum} for simplicity.
However, the eigenstates of the S-wave quarkonia can be also mixed by the magnetic fields, where the spin-singlet and spin-triplet with $S_z=0$ are mixed with each other.
Such a mixing effect between the S-wave quarkonia under a magnetic field is well-known by the constituent quark models \cite{Alford:2013jva,Bonati:2015dka,Suzuki:2016kcs,Yoshida:2016xgm}, effective Lagrangian \cite{Cho:2014exa,Cho:2014loa,Yoshida:2016xgm}, and QCD sum rules \cite{Cho:2014exa,Cho:2014loa}.
In particular, in the strong-field limit, the mixed configuration of S-wave quarkonia becomes $\frac{1}{\sqrt{2}}[Y_{00} \chi_{00} \pm Y_{00} \chi_{10}]$, while the components with $S_z=\pm1$ ($Y_{00} \chi_{11}$ and $Y_{00} \chi_{1-1}$) are not mixed.
As a result, the E1 decay properties from the P wave in the PB limit to the S wave in the strong-field limit are shown in Table \ref{Tab_decay_in_PB2_E1}.
From this table, we find that the decay width from each eigenstate will never change even if we consider the mixing between the final states.

\subsection{M1 decays in the PB limit} \label{Sec_M1}
M1 transitions change the spin angular momentum of quarkonia by $\Delta S = \pm 1$, while they conserve the orbital angular momentum, $\Delta L=0$.
In vacuum, the M1 decay processes, such as $J/\psi \to \eta_c \gamma$, $\chi_{c1}(\chi_{c2}) \to h_c \gamma$, and $h_c \to \chi_{c0} \gamma$ are possible.
Here, we focus on decays from P-wave quarkonia in the PB limit.

The M1 transition operator at leading order of $1/m$ is \cite{Brambilla:2005zw}
\begin{equation}
\mathcal{L}_\mathrm{M1}^{(0)} = \int d^3 r \mathrm{Tr} \left[ \frac{1}{2m} V_S^{(\sigma \cdot B)/m} \{\mathrm{S}^\dag, {\boldsymbol \sigma} \cdot ee_Q  {\bf B}^\mathrm{em} \} \mathrm{S} \right],
\end{equation}
where $V_S^{(\sigma \cdot B)/m} =1$ at the leading order of $1/m$, and ${\bf B}^\mathrm{em}$ is the external magnetic field.
Using this Lagrangian, the scattering amplitude is
\begin{eqnarray}
&& \mathcal{A}_\mathrm{M1}^{(0)} (2\pi)^3 \delta^3 ({\bf P}^\prime + {\bf k}) \nonumber\\
&& = - \langle H^\prime ( {\bf P}^\prime, \lambda^\prime ) \gamma ( {\bf k}, \sigma) | \int d^3R \, \mathcal{L}_{\mathrm{M1}}^{(0)} | H ( {\bf 0}, \lambda ) \rangle.
\end{eqnarray}
We use the external magnetic field projection on the one photon state:
\begin{equation}
\langle \gamma({\bf k}, \sigma) | {\bf B}^\mathrm{em} | \mathrm{vac} \rangle = -i {\bf k} \times {\boldsymbol \epsilon}^{\ast} (\sigma) e^{-i {\bf k} \cdot {\bf R}}.
\end{equation}

For the decays from the $S_z=0$ basis [(\ref{pol_base1}), (\ref{pol_base2}), (\ref{pol_base5}), and (\ref{pol_base6})], the amplitude squared is
\begin{eqnarray}
\sum_\sigma |\mathcal{A}_{\ket{L=1;10} \to \ket{L=1;00} \gamma} |^2 &=& \delta_{n n^\prime} \frac{4\pi \alpha_{em} e_Q^2 k^2}{m^2} \sin^2 \alpha. \nonumber \\
\end{eqnarray}
For the $S_z=\pm1$ basis [(\ref{pol_base3}), (\ref{pol_base4}), (\ref{pol_base7}), and (\ref{pol_base8})],
\begin{eqnarray}
\sum_\sigma |\mathcal{A}_{\ket{L=1;1\pm1} \to \ket{L=1;00} \gamma} |^2 &=& \delta_{n n^\prime} \frac{2\pi \alpha_{em} e_Q^2 k^2}{m^2}  ( \cos \alpha \pm 1)^2. \nonumber \\
\end{eqnarray}
Here we used the overlap integral for the spatial wave functions: $\int_0^\infty dr r^2 R_{nl} (r) R_{n^\prime l} (r) =\delta_{nn^\prime}$.

\begingroup
\renewcommand{\arraystretch}{1.5}
\begin{table*}[tbh!]
\centering
\caption{Summary of M1 decays between P-wave quarkonia in the PB limit.
$\alpha=0$ ($\alpha=\pi/2$) corresponds to the direction parallel (perpendicular) to the magnetic field.}
\begin{tabular}{ccccc}
\hline\hline
$\Delta J_z$ & Decay process & Decay factor & $\alpha=0,\pi$ & $\alpha=\pi/2$ \\
\hline
$0 \to +1$  & $ \frac{1}{\sqrt{2}} [Y_{10}  \chi_{00} \pm Y_{10} \chi_{10}] \to Y_{10} \chi_{11} $ & \large $\frac{1}{2} ( \cos^2 \alpha + 1)$ & $1$ & \large $\frac{1}{2}$  \\
$0 \to 0$  & $ \frac{1}{\sqrt{2}} [Y_{10}  \chi_{00} + Y_{10} \chi_{10}] \to \frac{1}{\sqrt{2}} [Y_{10}  \chi_{00} - Y_{10} \chi_{10}] $ & \large $ \sin^2 \alpha$ & $0$ & $1$  \\
$0 \to 0$  & $ \frac{1}{\sqrt{2}} [Y_{10}  \chi_{00} - Y_{10} \chi_{10}] \to \frac{1}{\sqrt{2}} [Y_{10}  \chi_{00} + Y_{10} \chi_{10}] $ & \large $ \sin^2 \alpha$ & $0$ & $1$  \\
$0 \to -1$  & $ \frac{1}{\sqrt{2}} [Y_{10}  \chi_{00} \pm Y_{10} \chi_{10}] \to Y_{10} \chi_{1-1} $ & \large $\frac{1}{2} ( \cos^2 \alpha + 1)$ & $1$ & \large $\frac{1}{2}$  \\
\hline
$0 \to +1 $  & $ \frac{1}{\sqrt{2}} [Y_{11} \chi_{1-1} \pm Y_{1-1} \chi_{11}] \to \frac{1}{\sqrt{2}} [Y_{11}  \chi_{00} \pm  Y_{11} \chi_{10}] $ & \large $\frac{1}{4} ( \cos^2 \alpha + 1)$ & \large $\frac{1}{2}$ & \large $\frac{1}{4}$ \\
$0 \to -1 $  & $ \frac{1}{\sqrt{2}} [Y_{11} \chi_{1-1} \pm Y_{1-1} \chi_{11}] \to \frac{1}{\sqrt{2}} [Y_{1-1}  \chi_{00} \pm  Y_{1-1} \chi_{10}] $ & \large $\frac{1}{4} ( \cos^2 \alpha + 1)$ & \large $\frac{1}{2}$ & \large $\frac{1}{4}$ \\
\hline \hline
$+1 \to +2$ & $ \frac{1}{\sqrt{2}} [Y_{11}  \chi_{00} \pm  Y_{11} \chi_{10}] \to Y_{1 1} \chi_{11} $ & \large $\frac{1}{2} ( \cos^2 \alpha + 1)$ & $1$ & \large $\frac{1}{2}$ \\
$+1 \to 0$ & $ \frac{1}{\sqrt{2}} [Y_{11}  \chi_{00} \pm  Y_{11} \chi_{10}] \to \frac{1}{\sqrt{2}} [Y_{11} \chi_{1-1} \pm Y_{1-1} \chi_{11}] $ & \large  $\frac{1}{4} ( \cos^2 \alpha + 1)$ & \large $\frac{1}{2}$ & \large $\frac{1}{4}$ \\
$\pm1 \to \pm1$ & $ \frac{1}{\sqrt{2}} [Y_{1\pm1}  \chi_{00} +  Y_{1\pm1} \chi_{10}] \to \frac{1}{\sqrt{2}} [Y_{1\pm1}  \chi_{00} - Y_{1\pm1} \chi_{10}] $ & \large $\sin^2 \alpha$ & $0$ & $1$ \\
$\pm1 \to \pm1$ & $ \frac{1}{\sqrt{2}} [Y_{1\pm1}  \chi_{00} -  Y_{1\pm1} \chi_{10}] \to \frac{1}{\sqrt{2}} [Y_{1\pm1}  \chi_{00} + Y_{1\pm1} \chi_{10}] $ & \large $\sin^2 \alpha$ & $0$ & $1$ \\
$-1 \to 0$ & $ \frac{1}{\sqrt{2}} [Y_{1-1}  \chi_{00} \pm  Y_{1-1} \chi_{10}] \to \frac{1}{\sqrt{2}} [Y_{11} \chi_{1-1} \pm Y_{1-1} \chi_{11}] $ & \large $\frac{1}{4} ( \cos^2 \alpha+ 1)$ & \large $\frac{1}{2}$ & \large $\frac{1}{4}$ \\
$-1 \to -2$ & $ \frac{1}{\sqrt{2}} [Y_{1-1}  \chi_{00} \pm  Y_{1-1} \chi_{10}] \to Y_{1 -1} \chi_{1-1} $ & \large $\frac{1}{2} ( \cos^2 \alpha + 1)$ & $1$ & \large $\frac{1}{2}$ \\
\hline
$+1 \to 0$ & $ Y_{10} \chi_{11} \to \frac{1}{\sqrt{2}} [Y_{10}  \chi_{00} \pm Y_{10} \chi_{10}] $ & \large $\frac{1}{2} ( \cos^2 \alpha + 1)$ & $1$ & \large $\frac{1}{2}$ \\
$-1 \to 0$ & $ Y_{10} \chi_{1-1} \to \frac{1}{\sqrt{2}} [Y_{10}  \chi_{00} \pm Y_{10} \chi_{10}] $ & \large $\frac{1}{2} ( \cos^2 \alpha + 1)$ & $1$ & \large $\frac{1}{2}$ \\
\hline \hline
$+2 \to +1$ & $ Y_{1 1} \chi_{11} \to \frac{1}{\sqrt{2}} [Y_{11}  \chi_{00} \pm Y_{11} \chi_{10}] $ & \large $\frac{1}{2} ( \cos^2 \alpha + 1)$ & $1$ & \large $\frac{1}{2}$ \\
$-2 \to -1$ & $ Y_{1 -1} \chi_{1-1} \to \frac{1}{\sqrt{2}} [Y_{1-1}  \chi_{00} \pm Y_{1-1} \chi_{10}] $ & \large $ \frac{1}{2}  ( \cos^2 \alpha + 1)$ & $1$ & \large $\frac{1}{2}$ \\
\hline\hline
\end{tabular}
\label{Tab_decay_in_PB_M1}
\end{table*}
\endgroup

Finally, we obtain the analytic formulas for the M1 decay widths:
\begin{widetext}
\begin{eqnarray}
\Gamma_{\ket{L=1;10} \to \ket{L=1;00} \gamma} &=& \frac{1}{8\pi^2} \left( 1-\frac{k_\gamma}{M_H} \right) \int_0^\infty dk k \int d \Omega(\hat{\bf{k}}) \delta(k-k_\gamma) \sum_\sigma |\mathcal{A}_{\ket{L=1;10} \to \ket{L=1;00} \gamma} |^2  \nonumber\\
&=& \delta_{n n^\prime} 2\alpha_{em} e_Q^2 \frac{1}{m^2} k_\gamma^3  \left(1- \frac{k_\gamma}{M_H} \right)  \sin^2 \alpha,  \label{PB_M1_width1}
\end{eqnarray}
\begin{eqnarray}
\Gamma_{\ket{L=1;1\pm1} \to \ket{L=1;00} \gamma} &=& \frac{1}{8\pi^2} \left( 1-\frac{k_\gamma}{M_H} \right) \int_0^\infty dk k \int d \Omega(\hat{{\bf k}}) \delta(k-k_\gamma) \sum_\sigma |\mathcal{A}_{\ket{L=1;1\pm1} \to \ket{L=1;00} \gamma} |^2  \nonumber\\
&=& \delta_{n n^\prime} \alpha_{em} e_Q^2 \frac{1}{m^2} k_\gamma^3 \left(1- \frac{k_\gamma}{M_H} \right) ( \cos^2 \alpha + 1).  \label{PB_M1_width2}
\end{eqnarray}
\end{widetext}

Note that the widths of the M1 decays from spin-triplet to singlet in vacuum ($^3 \! P_0, ^3 \! P_1, ^3 \! P_2 \to ^1 \! P_1$) have the factor of $4/3$, which have no angular dependence, as shown in Appendix \ref{App_vac}.

From these formulas, our findings are as follows:
\begin{itemize}
\item[(a)] By the factor of $\delta_{n n^\prime}$, only the transition between states with the same $n$ is allowed, which is unlike the E1 decays.
The (so-called hindered) transitions with $n \neq n^\prime$ appear as the relativistic corrections.
\item[(b)] The factors of $ \sin^2 \alpha$ and $\cos^2 \alpha + 1$ indicate the angular dependence of the decay width, which is the same as E1 decays.
\item[(c)] The M1 decay width from the {\it $S_z$-conserved} transition between $\ket{SS_z=10}$ and $\ket{SS_z=00}$ for $\alpha=\pi/2$ has the factor of $2$, which is larger than that in vacuum, so that the decays perpendicular to the magnetic field are enhanced, while the decay of $\ket{SS_z=10}$ for $\alpha=0,\pi$ is forbidden.
\item[(d)] The M1 decay widths from the {\it $S_z$-changed} transition between $\ket{SS_z=1\pm1}$ and $\ket{SS_z=00}$ for $\alpha=0,\pi$ has the factor of $2$, which is larger than that in vacuum, so that the decays parallel to the magnetic field is enhanced.
The decay width of $\ket{SS_z=1\pm1}$ for $\alpha=\pi/2$ has the factor of $1$, and the decays are suppressed.
\end{itemize}

By using the formulas, Eqs.~(\ref{PB_M1_width1}) and (\ref{PB_M1_width2}), and the eigenstates defined in Table \ref{Tab_CG_PB}, we can discuss the decay properties in the PB limit.
The M1 decay properties in the PB limit are summarized in Table \ref{Tab_decay_in_PB_M1}.
Here, we listed all of the allowed processes, but in the more realistic situation, we have to consider the possible processes within the phase space in momentum space.
Since $L_z$ of quarkonia is conserved for the E1 transitions, we classified the decay processes by the differences $\Delta J_z (=\Delta S_z)$ of quarkonia.

The $S_z(J_z)$-conserved transitions have the angle factor of $\sin^2 \alpha$.
This process is possible for the $J_z=0$ and $\pm1$, but it is forbidden for $J_z=\pm2$.
The $S_z(J_z)$-changed transitions have the angle factor of $ \cos^2 \alpha + 1$, and they can be divided into two types: the factors of $\frac{1}{2}$ and $\frac{1}{4}$.
The decays with $\frac{1}{2} (\cos^2 \alpha + 1)$ are caused by the pure state, such as $Y_{10} \chi_{1\pm1}$ (the middle state for $J_z=\pm1$) and $Y_{1 \pm1} \chi_{1 \pm1}$ ($J_z=\pm2$), in which there is no mixing between $Y_{LL_z} \chi_{SS_z}$ bases in the PB limit.
On the other hand, the decays with $\frac{1}{4} (\cos^2 \alpha + 1)$ are transitions between ``mixing" states.

\section{Conclusion and outlook} \label{Sec_conc}
In this work, we have investigated anisotropic radiative decays from quarkonia in the PB limit.
To this end, we have first developed the pNRQCD formalism with the polarized wave functions (\ref{pol_base1})--(\ref{pol_base8}).
Our main results are the analytic formulas for the E1 [Eqs.~(\ref{PB_E1_width1}), (\ref{PB_E1_width2}), (\ref{PB_E1_width3}) and (\ref{PB_E1_width4})] and M1 [Eqs.~(\ref{PB_M1_width1}) and (\ref{PB_M1_width2})] decay widths in the PB and nonrelativistic limit.
From these formulas, the properties of anisotropic decays are summarized in Tables \ref{Tab_decay_in_PB_E1}, \ref{Tab_decay_in_PB2_E1}, and \ref{Tab_decay_in_PB_M1}.

One of the advantages of pNRQCD is to introduce the higher-order corrections in a systematic way.
In fact, the E1 and M1 decay widths in the nonrelativistic limit (or at ``leading order") are of the order $k_\gamma^3/(m^2v^2)$ and $k_\gamma^3/m^2$, respectively.
The formulas shown in this work include the term with $k_\gamma/M_H \sim v^2$ as a correction, and this contribution is expected to be relatively suppressed, compared with the leading term.
Therefore, to include all of the higher order corrections at the relative order $v^2$ in the decay widths, we have to evaluate the corrections, such as higher-order transition operators, higher-order potentials, and higher-Fock states including the color-octet states.
Although the expansion parameters are $v^2 \sim 0.1$ for bottomonia and $v^2 \sim 0.3$ for charmonia, such corrections might be crucial for the quantitative estimate of the decay widths (see Refs.~\cite{Brambilla:2012be,Steinbeisser:2017hke,Segovia:2017nla} for the E1 transitions and Refs.~\cite{Brambilla:2005zw,Pineda:2013lta} for the M1 transitions).
Inclusion of the higher-order corrections into our formalism is left for future work.

As another approach to evaluate decay widths, we can use the constituent quark model in a magnetic field (\cite{Alford:2013jva,Bonati:2015dka,Suzuki:2016kcs,Yoshida:2016xgm} for the S wave and \cite{Bonati:2015dka, Iwasaki:2018pby} for the P wave), and it enables us to more quantitatively estimate radiative decay widths.
In particular, the investigation of quarkonia in a magnetic field will be useful for understanding the anisotropy of the confinement potential under strong magnetic fields, as estimated by phenomenological models \cite{Miransky:2002rp,Andreichikov:2012xe,Chernodub:2014uua,Rougemont:2014efa,Simonov:2015yka,Dudal:2016joz,Hasan:2017fmf,Singh:2017nfa} as well as lattice QCD simulations at zero \cite{Bonati:2014ksa} and finite temperature \cite{Bonati:2016kxj,Bonati:2017uvz}.
The relationship between such deformed potentials and decay properties will be also interesting.

\section*{Acknowledgments}
The authors thank Su Houng Lee for discussion in the early stage of this research, and we are grateful to Makoto Oka for giving us helpful comments on the manuscript.
This work is partially supported by the Grant-in-Aid for Scientific Research (Grant No.~17K14277) from the Japan Society for the Promotion of Science.
K. S. is supported by MEXT as ``Priority Issue on Post-K computer" (Elucidation of the Fundamental Laws and Evolution of the Universe) and Joint Institute for Computational Fundamental Science (JICFuS).

\appendix
\section{Decay widths in vacuum} \label{App_vac}
In the usual manner \cite{Brambilla:2005zw,Brambilla:2012be}, the radiative decay widths in vacuum can be calculated by using the wave function bases, Eqs.~(\ref{base_1S0})--(\ref{base_3P2}), where the scattering amplitude is summed over all the polarization of the photon and the initial and final quarkonia ($\sigma$, $\lambda$, and $\lambda^\prime$).
As another approach, we can reproduce the decay widths in vacuum by using the polarized bases (\ref{pol_base1})--(\ref{pol_base8}) introduced in this work.
By combining the polarized bases (\ref{pol_base1})--(\ref{pol_base8}) and the Clebsch-Gordan coefficients summarized in Table \ref{Tab_CG_vac}, we can reconstruct the wave functions of the P-wave eigenstates in vacuum.
Then, we sum over the photon polarization $\sigma$ and $J_z$ for the quarkonium.
As a result, for the arbitrary angular dependence $\alpha$, we can prove that the polarized bases can lead to the same as the decay widths from the usual bases.
Below, we summarize the results of the E1 and M1 decay widths.

\subsection{E1 decay width in vacuum}
The E1 decay width for both the singlet-to-singlet ($n^1 \! P_1 \to n^{\prime 1} \! S_0 \gamma$) and triplet-to-triplet ($n^3 \! P_J \to n^{\prime 3} \! S_1 \gamma$) transitions is \cite{Brambilla:2012be}
\begin{eqnarray}
\Gamma_{nP \to n^\prime S \gamma} &=& \frac{1}{8\pi^2} \left( 1-\frac{k_\gamma}{M_{nP}} \right) \int_0^\infty dk k \int d \Omega(\hat{ {\bf k}}) \delta(k-k_\gamma) \nonumber\\
 && \times \frac{1}{N_\lambda} \sum_{\lambda \lambda^\prime \sigma} |\mathcal{A}_{nP \to n^\prime S \gamma} |^2 \nonumber\\
&=& \frac{4}{9} \alpha_{em} e_Q^2 k_\gamma^3 \left[ \int_0^\infty dr r^3 R_{n^\prime L^\prime} (r) R_{nL} (r) \right]^2 \nonumber\\
 && \times\left(1- \frac{k_\gamma}{M_{nP}} \right).
\end{eqnarray}

\begin{widetext}
\subsection{M1 decay width in vacuum}
The M1 decay width for the triplet-to-singlet ($n^3 P_J \to n^{\prime 1} \! P_1 \gamma$) transition is \cite{Brambilla:2005zw}
\begin{eqnarray}
\Gamma_{n^3 P_J \to  \! n^{\prime 1} P_1 \! \gamma} &=& \frac{1}{8\pi^2} \left( 1-\frac{k_\gamma}{M_{n^3 P_J}} \right) \int_0^\infty dk k \int d \Omega(\hat{{\bf k}}) \delta(k-k_\gamma) \frac{1}{N_\lambda} \sum_{\lambda \lambda^\prime \sigma} |\mathcal{A}_{n^3 P_J \to n^{\prime 1} \! P_1 \gamma} |^2  \nonumber\\
&=& \delta_{n n^\prime} \frac{4}{3} \alpha_{em} e_Q^2 \frac{1}{m^2} k_\gamma^3  \left(1- \frac{k_\gamma}{M_{n^3 P_J}} \right).
\end{eqnarray}
For the singlet-to-triplet ($n^1 \! P_1 \to n^{\prime 3} P_J \gamma$) transition \cite{Brambilla:2005zw}
\begin{eqnarray}
\Gamma_{n^1 \! P_1 \to n^{\prime 3} P_J \gamma} &=& \frac{2J+1}{3} \delta_{n n^\prime} \frac{4}{3} \alpha_{em} e_Q^2 \frac{1}{m^2} k_\gamma^3  \left(1- \frac{k_\gamma}{M_{n^1 \! P_1}} \right).
\end{eqnarray}
\end{widetext}

\bibliography{HPBEdecay}

\begin{thebibliography}{67}%
\makeatletter
\providecommand \@ifxundefined [1]{%
 \@ifx{#1\undefined}
}%
\providecommand \@ifnum [1]{%
 \ifnum #1\expandafter \@firstoftwo
 \else \expandafter \@secondoftwo
 \fi
}%
\providecommand \@ifx [1]{%
 \ifx #1\expandafter \@firstoftwo
 \else \expandafter \@secondoftwo
 \fi
}%
\providecommand \natexlab [1]{#1}%
\providecommand \enquote  [1]{``#1''}%
\providecommand \bibnamefont  [1]{#1}%
\providecommand \bibfnamefont [1]{#1}%
\providecommand \citenamefont [1]{#1}%
\providecommand \href@noop [0]{\@secondoftwo}%
\providecommand \href [0]{\begingroup \@sanitize@url \@href}%
\providecommand \@href[1]{\@@startlink{#1}\@@href}%
\providecommand \@@href[1]{\endgroup#1\@@endlink}%
\providecommand \@sanitize@url [0]{\catcode `\\12\catcode `\$12\catcode
  `\&12\catcode `\#12\catcode `\^12\catcode `\_12\catcode `\%12\relax}%
\providecommand \@@startlink[1]{}%
\providecommand \@@endlink[0]{}%
\providecommand \url  [0]{\begingroup\@sanitize@url \@url }%
\providecommand \@url [1]{\endgroup\@href {#1}{\urlprefix }}%
\providecommand \urlprefix  [0]{URL }%
\providecommand \Eprint [0]{\href }%
\providecommand \doibase [0]{http://dx.doi.org/}%
\providecommand \selectlanguage [0]{\@gobble}%
\providecommand \bibinfo  [0]{\@secondoftwo}%
\providecommand \bibfield  [0]{\@secondoftwo}%
\providecommand \translation [1]{[#1]}%
\providecommand \BibitemOpen [0]{}%
\providecommand \bibitemStop [0]{}%
\providecommand \bibitemNoStop [0]{.\EOS\space}%
\providecommand \EOS [0]{\spacefactor3000\relax}%
\providecommand \BibitemShut  [1]{\csname bibitem#1\endcsname}%
\let\auto@bib@innerbib\@empty
\bibitem [{\citenamefont {Paschen}\ and\ \citenamefont
  {Back}(1921)}]{Paschen1921}%
  \BibitemOpen
  \bibfield  {author} {\bibinfo {author} {\bibfnamefont {F.}~\bibnamefont
  {Paschen}}\ and\ \bibinfo {author} {\bibfnamefont {E.}~\bibnamefont {Back}},\
  }\href@noop {} {\bibfield  {journal} {\bibinfo  {journal} {Physica}\ }\textbf
  {\bibinfo {volume} {1}},\ \bibinfo {pages} {261} (\bibinfo {year}
  {1921})}\BibitemShut {NoStop}%
\bibitem [{\citenamefont {Iwasaki}\ \emph {et~al.}(2018)\citenamefont
  {Iwasaki}, \citenamefont {Oka}, \citenamefont {Suzuki},\ and\ \citenamefont
  {Yoshida}}]{Iwasaki:2018pby}%
  \BibitemOpen
  \bibfield  {author} {\bibinfo {author} {\bibfnamefont {S.}~\bibnamefont
  {Iwasaki}}, \bibinfo {author} {\bibfnamefont {M.}~\bibnamefont {Oka}},
  \bibinfo {author} {\bibfnamefont {K.}~\bibnamefont {Suzuki}}, \ and\ \bibinfo
  {author} {\bibfnamefont {T.}~\bibnamefont {Yoshida}},\ }\href@noop {} {\
  (\bibinfo {year} {2018})},\ \Eprint {http://arxiv.org/abs/1802.04971}
  {arXiv:1802.04971 [hep-ph]} \BibitemShut {NoStop}%
\bibitem [{\citenamefont {Kharzeev}\ \emph {et~al.}(2008)\citenamefont
  {Kharzeev}, \citenamefont {McLerran},\ and\ \citenamefont
  {Warringa}}]{Kharzeev:2007jp}%
  \BibitemOpen
  \bibfield  {author} {\bibinfo {author} {\bibfnamefont {D.~E.}\ \bibnamefont
  {Kharzeev}}, \bibinfo {author} {\bibfnamefont {L.~D.}\ \bibnamefont
  {McLerran}}, \ and\ \bibinfo {author} {\bibfnamefont {H.~J.}\ \bibnamefont
  {Warringa}},\ }\href {\doibase 10.1016/j.nuclphysa.2008.02.298} {\bibfield
  {journal} {\bibinfo  {journal} {Nucl. Phys.}\ }\textbf {\bibinfo {volume}
  {A803}},\ \bibinfo {pages} {227} (\bibinfo {year} {2008})},\ \Eprint
  {http://arxiv.org/abs/0711.0950} {arXiv:0711.0950 [hep-ph]} \BibitemShut
  {NoStop}%
\bibitem [{\citenamefont {Skokov}\ \emph {et~al.}(2009)\citenamefont {Skokov},
  \citenamefont {Illarionov},\ and\ \citenamefont {Toneev}}]{Skokov:2009qp}%
  \BibitemOpen
  \bibfield  {author} {\bibinfo {author} {\bibfnamefont {V.}~\bibnamefont
  {Skokov}}, \bibinfo {author} {\bibfnamefont {A.~{\relax Yu}.}\ \bibnamefont
  {Illarionov}}, \ and\ \bibinfo {author} {\bibfnamefont {V.}~\bibnamefont
  {Toneev}},\ }\href {\doibase 10.1142/S0217751X09047570} {\bibfield  {journal}
  {\bibinfo  {journal} {Int. J. Mod. Phys.}\ }\textbf {\bibinfo {volume}
  {A24}},\ \bibinfo {pages} {5925} (\bibinfo {year} {2009})},\ \Eprint
  {http://arxiv.org/abs/0907.1396} {arXiv:0907.1396 [nucl-th]} \BibitemShut
  {NoStop}%
\bibitem [{\citenamefont {Voronyuk}\ \emph {et~al.}(2011)\citenamefont
  {Voronyuk}, \citenamefont {Toneev}, \citenamefont {Cassing}, \citenamefont
  {Bratkovskaya}, \citenamefont {Konchakovski},\ and\ \citenamefont
  {Voloshin}}]{Voronyuk:2011jd}%
  \BibitemOpen
  \bibfield  {author} {\bibinfo {author} {\bibfnamefont {V.}~\bibnamefont
  {Voronyuk}}, \bibinfo {author} {\bibfnamefont {V.~D.}\ \bibnamefont
  {Toneev}}, \bibinfo {author} {\bibfnamefont {W.}~\bibnamefont {Cassing}},
  \bibinfo {author} {\bibfnamefont {E.~L.}\ \bibnamefont {Bratkovskaya}},
  \bibinfo {author} {\bibfnamefont {V.~P.}\ \bibnamefont {Konchakovski}}, \
  and\ \bibinfo {author} {\bibfnamefont {S.~A.}\ \bibnamefont {Voloshin}},\
  }\href {\doibase 10.1103/PhysRevC.83.054911} {\bibfield  {journal} {\bibinfo
  {journal} {Phys. Rev.}\ }\textbf {\bibinfo {volume} {C83}},\ \bibinfo {pages}
  {054911} (\bibinfo {year} {2011})},\ \Eprint {http://arxiv.org/abs/1103.4239}
  {arXiv:1103.4239 [nucl-th]} \BibitemShut {NoStop}%
\bibitem [{\citenamefont {Ou}\ and\ \citenamefont {Li}(2011)}]{Ou:2011fm}%
  \BibitemOpen
  \bibfield  {author} {\bibinfo {author} {\bibfnamefont {L.}~\bibnamefont
  {Ou}}\ and\ \bibinfo {author} {\bibfnamefont {B.-A.}\ \bibnamefont {Li}},\
  }\href {\doibase 10.1103/PhysRevC.84.064605} {\bibfield  {journal} {\bibinfo
  {journal} {Phys. Rev.}\ }\textbf {\bibinfo {volume} {C84}},\ \bibinfo {pages}
  {064605} (\bibinfo {year} {2011})},\ \Eprint {http://arxiv.org/abs/1107.3192}
  {arXiv:1107.3192 [nucl-th]} \BibitemShut {NoStop}%
\bibitem [{\citenamefont {Bzdak}\ and\ \citenamefont
  {Skokov}(2012)}]{Bzdak:2011yy}%
  \BibitemOpen
  \bibfield  {author} {\bibinfo {author} {\bibfnamefont {A.}~\bibnamefont
  {Bzdak}}\ and\ \bibinfo {author} {\bibfnamefont {V.}~\bibnamefont {Skokov}},\
  }\href {\doibase 10.1016/j.physletb.2012.02.065} {\bibfield  {journal}
  {\bibinfo  {journal} {Phys. Lett.}\ }\textbf {\bibinfo {volume} {B710}},\
  \bibinfo {pages} {171} (\bibinfo {year} {2012})},\ \Eprint
  {http://arxiv.org/abs/1111.1949} {arXiv:1111.1949 [hep-ph]} \BibitemShut
  {NoStop}%
\bibitem [{\citenamefont {Deng}\ and\ \citenamefont
  {Huang}(2012)}]{Deng:2012pc}%
  \BibitemOpen
  \bibfield  {author} {\bibinfo {author} {\bibfnamefont {W.-T.}\ \bibnamefont
  {Deng}}\ and\ \bibinfo {author} {\bibfnamefont {X.-G.}\ \bibnamefont
  {Huang}},\ }\href {\doibase 10.1103/PhysRevC.85.044907} {\bibfield  {journal}
  {\bibinfo  {journal} {Phys. Rev.}\ }\textbf {\bibinfo {volume} {C85}},\
  \bibinfo {pages} {044907} (\bibinfo {year} {2012})},\ \Eprint
  {http://arxiv.org/abs/1201.5108} {arXiv:1201.5108 [nucl-th]} \BibitemShut
  {NoStop}%
\bibitem [{\citenamefont {Bloczynski}\ \emph {et~al.}(2013)\citenamefont
  {Bloczynski}, \citenamefont {Huang}, \citenamefont {Zhang},\ and\
  \citenamefont {Liao}}]{Bloczynski:2012en}%
  \BibitemOpen
  \bibfield  {author} {\bibinfo {author} {\bibfnamefont {J.}~\bibnamefont
  {Bloczynski}}, \bibinfo {author} {\bibfnamefont {X.-G.}\ \bibnamefont
  {Huang}}, \bibinfo {author} {\bibfnamefont {X.}~\bibnamefont {Zhang}}, \ and\
  \bibinfo {author} {\bibfnamefont {J.}~\bibnamefont {Liao}},\ }\href {\doibase
  10.1016/j.physletb.2012.12.030} {\bibfield  {journal} {\bibinfo  {journal}
  {Phys. Lett.}\ }\textbf {\bibinfo {volume} {B718}},\ \bibinfo {pages} {1529}
  (\bibinfo {year} {2013})},\ \Eprint {http://arxiv.org/abs/1209.6594}
  {arXiv:1209.6594 [nucl-th]} \BibitemShut {NoStop}%
\bibitem [{\citenamefont {Bloczynski}\ \emph {et~al.}(2015)\citenamefont
  {Bloczynski}, \citenamefont {Huang}, \citenamefont {Zhang},\ and\
  \citenamefont {Liao}}]{Bloczynski:2013mca}%
  \BibitemOpen
  \bibfield  {author} {\bibinfo {author} {\bibfnamefont {J.}~\bibnamefont
  {Bloczynski}}, \bibinfo {author} {\bibfnamefont {X.-G.}\ \bibnamefont
  {Huang}}, \bibinfo {author} {\bibfnamefont {X.}~\bibnamefont {Zhang}}, \ and\
  \bibinfo {author} {\bibfnamefont {J.}~\bibnamefont {Liao}},\ }\href {\doibase
  10.1016/j.nuclphysa.2015.03.012} {\bibfield  {journal} {\bibinfo  {journal}
  {Nucl. Phys.}\ }\textbf {\bibinfo {volume} {A939}},\ \bibinfo {pages} {85}
  (\bibinfo {year} {2015})},\ \Eprint {http://arxiv.org/abs/1311.5451}
  {arXiv:1311.5451 [nucl-th]} \BibitemShut {NoStop}%
\bibitem [{\citenamefont {Deng}\ and\ \citenamefont
  {Huang}(2015)}]{Deng:2014uja}%
  \BibitemOpen
  \bibfield  {author} {\bibinfo {author} {\bibfnamefont {W.-T.}\ \bibnamefont
  {Deng}}\ and\ \bibinfo {author} {\bibfnamefont {X.-G.}\ \bibnamefont
  {Huang}},\ }\href {\doibase 10.1016/j.physletb.2015.01.050} {\bibfield
  {journal} {\bibinfo  {journal} {Phys. Lett.}\ }\textbf {\bibinfo {volume}
  {B742}},\ \bibinfo {pages} {296} (\bibinfo {year} {2015})},\ \Eprint
  {http://arxiv.org/abs/1411.2733} {arXiv:1411.2733 [nucl-th]} \BibitemShut
  {NoStop}%
\bibitem [{\citenamefont {Huang}(2016)}]{Huang:2015oca}%
  \BibitemOpen
  \bibfield  {author} {\bibinfo {author} {\bibfnamefont {X.-G.}\ \bibnamefont
  {Huang}},\ }\href {\doibase 10.1088/0034-4885/79/7/076302} {\bibfield
  {journal} {\bibinfo  {journal} {Rept. Prog. Phys.}\ }\textbf {\bibinfo
  {volume} {79}},\ \bibinfo {pages} {076302} (\bibinfo {year} {2016})},\
  \Eprint {http://arxiv.org/abs/1509.04073} {arXiv:1509.04073 [nucl-th]}
  \BibitemShut {NoStop}%
\bibitem [{\citenamefont {Hattori}\ and\ \citenamefont
  {Huang}(2017)}]{Hattori:2016emy}%
  \BibitemOpen
  \bibfield  {author} {\bibinfo {author} {\bibfnamefont {K.}~\bibnamefont
  {Hattori}}\ and\ \bibinfo {author} {\bibfnamefont {X.-G.}\ \bibnamefont
  {Huang}},\ }\href {\doibase 10.1007/s41365-016-0178-3} {\bibfield  {journal}
  {\bibinfo  {journal} {Nucl. Sci. Tech.}\ }\textbf {\bibinfo {volume} {28}},\
  \bibinfo {pages} {26} (\bibinfo {year} {2017})},\ \Eprint
  {http://arxiv.org/abs/1609.00747} {arXiv:1609.00747 [nucl-th]} \BibitemShut
  {NoStop}%
\bibitem [{\citenamefont {Zhao}\ \emph {et~al.}(2018)\citenamefont {Zhao},
  \citenamefont {Ma},\ and\ \citenamefont {Ma}}]{Zhao:2017rpf}%
  \BibitemOpen
  \bibfield  {author} {\bibinfo {author} {\bibfnamefont {X.-L.}\ \bibnamefont
  {Zhao}}, \bibinfo {author} {\bibfnamefont {Y.-G.}\ \bibnamefont {Ma}}, \ and\
  \bibinfo {author} {\bibfnamefont {G.-L.}\ \bibnamefont {Ma}},\ }\href
  {\doibase 10.1103/PhysRevC.97.024910} {\bibfield  {journal} {\bibinfo
  {journal} {Phys. Rev.}\ }\textbf {\bibinfo {volume} {C97}},\ \bibinfo {pages}
  {024910} (\bibinfo {year} {2018})},\ \Eprint
  {http://arxiv.org/abs/1709.05962} {arXiv:1709.05962 [hep-ph]} \BibitemShut
  {NoStop}%
\bibitem [{\citenamefont {Marasinghe}\ and\ \citenamefont
  {Tuchin}(2011)}]{Marasinghe:2011bt}%
  \BibitemOpen
  \bibfield  {author} {\bibinfo {author} {\bibfnamefont {K.}~\bibnamefont
  {Marasinghe}}\ and\ \bibinfo {author} {\bibfnamefont {K.}~\bibnamefont
  {Tuchin}},\ }\href {\doibase 10.1103/PhysRevC.84.044908} {\bibfield
  {journal} {\bibinfo  {journal} {Phys. Rev.}\ }\textbf {\bibinfo {volume}
  {C84}},\ \bibinfo {pages} {044908} (\bibinfo {year} {2011})},\ \Eprint
  {http://arxiv.org/abs/1103.1329} {arXiv:1103.1329 [hep-ph]} \BibitemShut
  {NoStop}%
\bibitem [{\citenamefont {Tuchin}(2011)}]{Tuchin:2011cg}%
  \BibitemOpen
  \bibfield  {author} {\bibinfo {author} {\bibfnamefont {K.}~\bibnamefont
  {Tuchin}},\ }\href {\doibase 10.1016/j.physletb.2011.10.047} {\bibfield
  {journal} {\bibinfo  {journal} {Phys. Lett.}\ }\textbf {\bibinfo {volume}
  {B705}},\ \bibinfo {pages} {482} (\bibinfo {year} {2011})},\ \Eprint
  {http://arxiv.org/abs/1105.5360} {arXiv:1105.5360 [nucl-th]} \BibitemShut
  {NoStop}%
\bibitem [{\citenamefont {Yang}\ and\ \citenamefont {M{\"
  u}ller}(2012)}]{Yang:2011cz}%
  \BibitemOpen
  \bibfield  {author} {\bibinfo {author} {\bibfnamefont {D.-L.}\ \bibnamefont
  {Yang}}\ and\ \bibinfo {author} {\bibfnamefont {B.}~\bibnamefont {M{\"
  u}ller}},\ }\href {\doibase 10.1088/0954-3899/39/1/015007} {\bibfield
  {journal} {\bibinfo  {journal} {J. Phys.}\ }\textbf {\bibinfo {volume}
  {G39}},\ \bibinfo {pages} {015007} (\bibinfo {year} {2012})},\ \Eprint
  {http://arxiv.org/abs/1108.2525} {arXiv:1108.2525 [hep-ph]} \BibitemShut
  {NoStop}%
\bibitem [{\citenamefont {Tuchin}(2013)}]{Tuchin:2013ie}%
  \BibitemOpen
  \bibfield  {author} {\bibinfo {author} {\bibfnamefont {K.}~\bibnamefont
  {Tuchin}},\ }\href {\doibase 10.1155/2013/490495} {\bibfield  {journal}
  {\bibinfo  {journal} {Adv. High Energy Phys.}\ }\textbf {\bibinfo {volume}
  {2013}},\ \bibinfo {pages} {490495} (\bibinfo {year} {2013})},\ \Eprint
  {http://arxiv.org/abs/1301.0099} {arXiv:1301.0099 [hep-ph]} \BibitemShut
  {NoStop}%
\bibitem [{\citenamefont {Machado}\ \emph {et~al.}(2013)\citenamefont
  {Machado}, \citenamefont {Navarra}, \citenamefont {de~Oliveira},
  \citenamefont {Noronha},\ and\ \citenamefont {Strickland}}]{Machado:2013rta}%
  \BibitemOpen
  \bibfield  {author} {\bibinfo {author} {\bibfnamefont {C.~S.}\ \bibnamefont
  {Machado}}, \bibinfo {author} {\bibfnamefont {F.~S.}\ \bibnamefont
  {Navarra}}, \bibinfo {author} {\bibfnamefont {E.~G.}\ \bibnamefont
  {de~Oliveira}}, \bibinfo {author} {\bibfnamefont {J.}~\bibnamefont
  {Noronha}}, \ and\ \bibinfo {author} {\bibfnamefont {M.}~\bibnamefont
  {Strickland}},\ }\href {\doibase 10.1103/PhysRevD.88.034009} {\bibfield
  {journal} {\bibinfo  {journal} {Phys. Rev.}\ }\textbf {\bibinfo {volume}
  {D88}},\ \bibinfo {pages} {034009} (\bibinfo {year} {2013})},\ \Eprint
  {http://arxiv.org/abs/1305.3308} {arXiv:1305.3308 [hep-ph]} \BibitemShut
  {NoStop}%
\bibitem [{\citenamefont {Alford}\ and\ \citenamefont
  {Strickland}(2013)}]{Alford:2013jva}%
  \BibitemOpen
  \bibfield  {author} {\bibinfo {author} {\bibfnamefont {J.}~\bibnamefont
  {Alford}}\ and\ \bibinfo {author} {\bibfnamefont {M.}~\bibnamefont
  {Strickland}},\ }\href {\doibase 10.1103/PhysRevD.88.105017} {\bibfield
  {journal} {\bibinfo  {journal} {Phys.Rev.}\ }\textbf {\bibinfo {volume}
  {D88}},\ \bibinfo {pages} {105017} (\bibinfo {year} {2013})},\ \Eprint
  {http://arxiv.org/abs/1309.3003} {arXiv:1309.3003 [hep-ph]} \BibitemShut
  {NoStop}%
\bibitem [{\citenamefont {Cho}\ \emph {et~al.}(2014)\citenamefont {Cho},
  \citenamefont {Hattori}, \citenamefont {Lee}, \citenamefont {Morita},\ and\
  \citenamefont {Ozaki}}]{Cho:2014exa}%
  \BibitemOpen
  \bibfield  {author} {\bibinfo {author} {\bibfnamefont {S.}~\bibnamefont
  {Cho}}, \bibinfo {author} {\bibfnamefont {K.}~\bibnamefont {Hattori}},
  \bibinfo {author} {\bibfnamefont {S.~H.}\ \bibnamefont {Lee}}, \bibinfo
  {author} {\bibfnamefont {K.}~\bibnamefont {Morita}}, \ and\ \bibinfo {author}
  {\bibfnamefont {S.}~\bibnamefont {Ozaki}},\ }\href {\doibase
  10.1103/PhysRevLett.113.172301} {\bibfield  {journal} {\bibinfo  {journal}
  {Phys. Rev. Lett.}\ }\textbf {\bibinfo {volume} {113}},\ \bibinfo {pages}
  {172301} (\bibinfo {year} {2014})},\ \Eprint {http://arxiv.org/abs/1406.4586}
  {arXiv:1406.4586 [hep-ph]} \BibitemShut {NoStop}%
\bibitem [{\citenamefont {Dudal}\ and\ \citenamefont
  {Mertens}(2015)}]{Dudal:2014jfa}%
  \BibitemOpen
  \bibfield  {author} {\bibinfo {author} {\bibfnamefont {D.}~\bibnamefont
  {Dudal}}\ and\ \bibinfo {author} {\bibfnamefont {T.~G.}\ \bibnamefont
  {Mertens}},\ }\href {\doibase 10.1103/PhysRevD.91.086002} {\bibfield
  {journal} {\bibinfo  {journal} {Phys. Rev.}\ }\textbf {\bibinfo {volume}
  {D91}},\ \bibinfo {pages} {086002} (\bibinfo {year} {2015})},\ \Eprint
  {http://arxiv.org/abs/1410.3297} {arXiv:1410.3297 [hep-th]} \BibitemShut
  {NoStop}%
\bibitem [{\citenamefont {Cho}\ \emph {et~al.}(2015)\citenamefont {Cho},
  \citenamefont {Hattori}, \citenamefont {Lee}, \citenamefont {Morita},\ and\
  \citenamefont {Ozaki}}]{Cho:2014loa}%
  \BibitemOpen
  \bibfield  {author} {\bibinfo {author} {\bibfnamefont {S.}~\bibnamefont
  {Cho}}, \bibinfo {author} {\bibfnamefont {K.}~\bibnamefont {Hattori}},
  \bibinfo {author} {\bibfnamefont {S.~H.}\ \bibnamefont {Lee}}, \bibinfo
  {author} {\bibfnamefont {K.}~\bibnamefont {Morita}}, \ and\ \bibinfo {author}
  {\bibfnamefont {S.}~\bibnamefont {Ozaki}},\ }\href {\doibase
  10.1103/PhysRevD.91.045025} {\bibfield  {journal} {\bibinfo  {journal} {Phys.
  Rev.}\ }\textbf {\bibinfo {volume} {D91}},\ \bibinfo {pages} {045025}
  (\bibinfo {year} {2015})},\ \Eprint {http://arxiv.org/abs/1411.7675}
  {arXiv:1411.7675 [hep-ph]} \BibitemShut {NoStop}%
\bibitem [{\citenamefont {Guo}\ \emph {et~al.}(2015)\citenamefont {Guo},
  \citenamefont {Shi}, \citenamefont {Xu}, \citenamefont {Xu},\ and\
  \citenamefont {Zhuang}}]{Guo:2015nsa}%
  \BibitemOpen
  \bibfield  {author} {\bibinfo {author} {\bibfnamefont {X.}~\bibnamefont
  {Guo}}, \bibinfo {author} {\bibfnamefont {S.}~\bibnamefont {Shi}}, \bibinfo
  {author} {\bibfnamefont {N.}~\bibnamefont {Xu}}, \bibinfo {author}
  {\bibfnamefont {Z.}~\bibnamefont {Xu}}, \ and\ \bibinfo {author}
  {\bibfnamefont {P.}~\bibnamefont {Zhuang}},\ }\href {\doibase
  10.1016/j.physletb.2015.10.038} {\bibfield  {journal} {\bibinfo  {journal}
  {Phys. Lett.}\ }\textbf {\bibinfo {volume} {B751}},\ \bibinfo {pages} {215}
  (\bibinfo {year} {2015})},\ \Eprint {http://arxiv.org/abs/1502.04407}
  {arXiv:1502.04407 [hep-ph]} \BibitemShut {NoStop}%
\bibitem [{\citenamefont {Bonati}\ \emph {et~al.}(2015)\citenamefont {Bonati},
  \citenamefont {D'Elia},\ and\ \citenamefont {Rucci}}]{Bonati:2015dka}%
  \BibitemOpen
  \bibfield  {author} {\bibinfo {author} {\bibfnamefont {C.}~\bibnamefont
  {Bonati}}, \bibinfo {author} {\bibfnamefont {M.}~\bibnamefont {D'Elia}}, \
  and\ \bibinfo {author} {\bibfnamefont {A.}~\bibnamefont {Rucci}},\ }\href
  {\doibase 10.1103/PhysRevD.92.054014} {\bibfield  {journal} {\bibinfo
  {journal} {Phys. Rev.}\ }\textbf {\bibinfo {volume} {D92}},\ \bibinfo {pages}
  {054014} (\bibinfo {year} {2015})},\ \Eprint
  {http://arxiv.org/abs/1506.07890} {arXiv:1506.07890 [hep-ph]} \BibitemShut
  {NoStop}%
\bibitem [{\citenamefont {Sadofyev}\ and\ \citenamefont
  {Yin}(2016{\natexlab{a}})}]{Sadofyev:2015hxa}%
  \BibitemOpen
  \bibfield  {author} {\bibinfo {author} {\bibfnamefont {A.~V.}\ \bibnamefont
  {Sadofyev}}\ and\ \bibinfo {author} {\bibfnamefont {Y.}~\bibnamefont {Yin}},\
  }\href {\doibase 10.1007/JHEP01(2016)052} {\bibfield  {journal} {\bibinfo
  {journal} {JHEP}\ }\textbf {\bibinfo {volume} {01}},\ \bibinfo {pages} {052}
  (\bibinfo {year} {2016}{\natexlab{a}})},\ \Eprint
  {http://arxiv.org/abs/1510.06760} {arXiv:1510.06760 [hep-th]} \BibitemShut
  {NoStop}%
\bibitem [{\citenamefont {Suzuki}\ and\ \citenamefont
  {Yoshida}(2016)}]{Suzuki:2016kcs}%
  \BibitemOpen
  \bibfield  {author} {\bibinfo {author} {\bibfnamefont {K.}~\bibnamefont
  {Suzuki}}\ and\ \bibinfo {author} {\bibfnamefont {T.}~\bibnamefont
  {Yoshida}},\ }\href {\doibase 10.1103/PhysRevD.93.051502} {\bibfield
  {journal} {\bibinfo  {journal} {Phys. Rev.}\ }\textbf {\bibinfo {volume}
  {D93}},\ \bibinfo {pages} {051502} (\bibinfo {year} {2016})},\ \Eprint
  {http://arxiv.org/abs/1601.02178} {arXiv:1601.02178 [hep-ph]} \BibitemShut
  {NoStop}%
\bibitem [{\citenamefont {Yoshida}\ and\ \citenamefont
  {Suzuki}(2016)}]{Yoshida:2016xgm}%
  \BibitemOpen
  \bibfield  {author} {\bibinfo {author} {\bibfnamefont {T.}~\bibnamefont
  {Yoshida}}\ and\ \bibinfo {author} {\bibfnamefont {K.}~\bibnamefont
  {Suzuki}},\ }\href {\doibase 10.1103/PhysRevD.94.074043} {\bibfield
  {journal} {\bibinfo  {journal} {Phys. Rev.}\ }\textbf {\bibinfo {volume}
  {D94}},\ \bibinfo {pages} {074043} (\bibinfo {year} {2016})},\ \Eprint
  {http://arxiv.org/abs/1607.04935} {arXiv:1607.04935 [hep-ph]} \BibitemShut
  {NoStop}%
\bibitem [{\citenamefont {Suzuki}\ and\ \citenamefont
  {Lee}(2017)}]{Suzuki:2016fof}%
  \BibitemOpen
  \bibfield  {author} {\bibinfo {author} {\bibfnamefont {K.}~\bibnamefont
  {Suzuki}}\ and\ \bibinfo {author} {\bibfnamefont {S.~H.}\ \bibnamefont
  {Lee}},\ }\href {\doibase 10.1103/PhysRevC.96.035203} {\bibfield  {journal}
  {\bibinfo  {journal} {Phys. Rev.}\ }\textbf {\bibinfo {volume} {C96}},\
  \bibinfo {pages} {035203} (\bibinfo {year} {2017})},\ \Eprint
  {http://arxiv.org/abs/1610.09853} {arXiv:1610.09853 [hep-ph]} \BibitemShut
  {NoStop}%
\bibitem [{\citenamefont {Hoelck}\ and\ \citenamefont
  {Wolschin}(2017)}]{Hoelck:2017dby}%
  \BibitemOpen
  \bibfield  {author} {\bibinfo {author} {\bibfnamefont {J.}~\bibnamefont
  {Hoelck}}\ and\ \bibinfo {author} {\bibfnamefont {G.}~\bibnamefont
  {Wolschin}},\ }\href {\doibase 10.1140/epja/i2017-12441-0} {\bibfield
  {journal} {\bibinfo  {journal} {Eur. Phys. J.}\ }\textbf {\bibinfo {volume}
  {A53}},\ \bibinfo {pages} {241} (\bibinfo {year} {2017})},\ \Eprint
  {http://arxiv.org/abs/1712.06871} {arXiv:1712.06871 [hep-ph]} \BibitemShut
  {NoStop}%
\bibitem [{\citenamefont {Braga}\ and\ \citenamefont
  {Ferreira}(2018)}]{Braga:2018zlu}%
  \BibitemOpen
  \bibfield  {author} {\bibinfo {author} {\bibfnamefont {N.~R.~F.}\
  \bibnamefont {Braga}}\ and\ \bibinfo {author} {\bibfnamefont {L.~F.}\
  \bibnamefont {Ferreira}},\ }\href {\doibase 10.1016/j.physletb.2018.06.053}
  {\bibfield  {journal} {\bibinfo  {journal} {Phys. Lett.}\ }\textbf {\bibinfo
  {volume} {B783}},\ \bibinfo {pages} {186} (\bibinfo {year} {2018})},\ \Eprint
  {http://arxiv.org/abs/1802.02084} {arXiv:1802.02084 [hep-ph]} \BibitemShut
  {NoStop}%
\bibitem [{\citenamefont {Machado}\ \emph {et~al.}(2014)\citenamefont
  {Machado}, \citenamefont {Matheus}, \citenamefont {Finazzo},\ and\
  \citenamefont {Noronha}}]{Machado:2013yaa}%
  \BibitemOpen
  \bibfield  {author} {\bibinfo {author} {\bibfnamefont {C.~S.}\ \bibnamefont
  {Machado}}, \bibinfo {author} {\bibfnamefont {R.~D.}\ \bibnamefont
  {Matheus}}, \bibinfo {author} {\bibfnamefont {S.~I.}\ \bibnamefont
  {Finazzo}}, \ and\ \bibinfo {author} {\bibfnamefont {J.}~\bibnamefont
  {Noronha}},\ }\href {\doibase 10.1103/PhysRevD.89.074027} {\bibfield
  {journal} {\bibinfo  {journal} {Phys. Rev.}\ }\textbf {\bibinfo {volume}
  {D89}},\ \bibinfo {pages} {074027} (\bibinfo {year} {2014})},\ \Eprint
  {http://arxiv.org/abs/1307.1797} {arXiv:1307.1797 [hep-ph]} \BibitemShut
  {NoStop}%
\bibitem [{\citenamefont {Gubler}\ \emph {et~al.}(2016)\citenamefont {Gubler},
  \citenamefont {Hattori}, \citenamefont {Lee}, \citenamefont {Oka},
  \citenamefont {Ozaki},\ and\ \citenamefont {Suzuki}}]{Gubler:2015qok}%
  \BibitemOpen
  \bibfield  {author} {\bibinfo {author} {\bibfnamefont {P.}~\bibnamefont
  {Gubler}}, \bibinfo {author} {\bibfnamefont {K.}~\bibnamefont {Hattori}},
  \bibinfo {author} {\bibfnamefont {S.~H.}\ \bibnamefont {Lee}}, \bibinfo
  {author} {\bibfnamefont {M.}~\bibnamefont {Oka}}, \bibinfo {author}
  {\bibfnamefont {S.}~\bibnamefont {Ozaki}}, \ and\ \bibinfo {author}
  {\bibfnamefont {K.}~\bibnamefont {Suzuki}},\ }\href {\doibase
  10.1103/PhysRevD.93.054026} {\bibfield  {journal} {\bibinfo  {journal} {Phys.
  Rev.}\ }\textbf {\bibinfo {volume} {D93}},\ \bibinfo {pages} {054026}
  (\bibinfo {year} {2016})},\ \Eprint {http://arxiv.org/abs/1512.08864}
  {arXiv:1512.08864 [hep-ph]} \BibitemShut {NoStop}%
\bibitem [{\citenamefont {Reddy~P.}\ \emph {et~al.}(2018)\citenamefont
  {Reddy~P.}, \citenamefont {Jahan C.~S.}, \citenamefont {Dhale}, \citenamefont
  {Mishra},\ and\ \citenamefont {Schaffner-Bielich}}]{Reddy:2017pqp}%
  \BibitemOpen
  \bibfield  {author} {\bibinfo {author} {\bibfnamefont {S.}~\bibnamefont
  {Reddy~P.}}, \bibinfo {author} {\bibfnamefont {A.}~\bibnamefont {Jahan
  C.~S.}}, \bibinfo {author} {\bibfnamefont {N.}~\bibnamefont {Dhale}},
  \bibinfo {author} {\bibfnamefont {A.}~\bibnamefont {Mishra}}, \ and\ \bibinfo
  {author} {\bibfnamefont {J.}~\bibnamefont {Schaffner-Bielich}},\ }\href
  {\doibase 10.1103/PhysRevC.97.065208} {\bibfield  {journal} {\bibinfo
  {journal} {Phys. Rev.}\ }\textbf {\bibinfo {volume} {C97}},\ \bibinfo {pages}
  {065208} (\bibinfo {year} {2018})},\ \Eprint
  {http://arxiv.org/abs/1712.07997} {arXiv:1712.07997 [nucl-th]} \BibitemShut
  {NoStop}%
\bibitem [{\citenamefont {Dhale}\ \emph {et~al.}(2018)\citenamefont {Dhale},
  \citenamefont {Reddy~P.}, \citenamefont {Jahan C.~S.},\ and\ \citenamefont
  {Mishra}}]{Dhale:2018plh}%
  \BibitemOpen
  \bibfield  {author} {\bibinfo {author} {\bibfnamefont {N.}~\bibnamefont
  {Dhale}}, \bibinfo {author} {\bibfnamefont {S.}~\bibnamefont {Reddy~P.}},
  \bibinfo {author} {\bibfnamefont {A.}~\bibnamefont {Jahan C.~S.}}, \ and\
  \bibinfo {author} {\bibfnamefont {A.}~\bibnamefont {Mishra}},\ }\href
  {\doibase 10.1103/PhysRevC.98.015202} {\bibfield  {journal} {\bibinfo
  {journal} {Phys. Rev.}\ }\textbf {\bibinfo {volume} {C98}},\ \bibinfo {pages}
  {015202} (\bibinfo {year} {2018})},\ \Eprint
  {http://arxiv.org/abs/1801.06405} {arXiv:1801.06405 [nucl-th]} \BibitemShut
  {NoStop}%
\bibitem [{\citenamefont {Fukushima}\ \emph {et~al.}(2016)\citenamefont
  {Fukushima}, \citenamefont {Hattori}, \citenamefont {Yee},\ and\
  \citenamefont {Yin}}]{Fukushima:2015wck}%
  \BibitemOpen
  \bibfield  {author} {\bibinfo {author} {\bibfnamefont {K.}~\bibnamefont
  {Fukushima}}, \bibinfo {author} {\bibfnamefont {K.}~\bibnamefont {Hattori}},
  \bibinfo {author} {\bibfnamefont {H.-U.}\ \bibnamefont {Yee}}, \ and\
  \bibinfo {author} {\bibfnamefont {Y.}~\bibnamefont {Yin}},\ }\href {\doibase
  10.1103/PhysRevD.93.074028} {\bibfield  {journal} {\bibinfo  {journal} {Phys.
  Rev.}\ }\textbf {\bibinfo {volume} {D93}},\ \bibinfo {pages} {074028}
  (\bibinfo {year} {2016})},\ \Eprint {http://arxiv.org/abs/1512.03689}
  {arXiv:1512.03689 [hep-ph]} \BibitemShut {NoStop}%
\bibitem [{\citenamefont {Finazzo}\ \emph {et~al.}(2016)\citenamefont
  {Finazzo}, \citenamefont {Critelli}, \citenamefont {Rougemont},\ and\
  \citenamefont {Noronha}}]{Finazzo:2016mhm}%
  \BibitemOpen
  \bibfield  {author} {\bibinfo {author} {\bibfnamefont {S.~I.}\ \bibnamefont
  {Finazzo}}, \bibinfo {author} {\bibfnamefont {R.}~\bibnamefont {Critelli}},
  \bibinfo {author} {\bibfnamefont {R.}~\bibnamefont {Rougemont}}, \ and\
  \bibinfo {author} {\bibfnamefont {J.}~\bibnamefont {Noronha}},\ }\href
  {\doibase 10.1103/PhysRevD.94.054020} {\bibfield  {journal} {\bibinfo
  {journal} {Phys. Rev.}\ }\textbf {\bibinfo {volume} {D94}},\ \bibinfo {pages}
  {054020} (\bibinfo {year} {2016})},\ \Eprint
  {http://arxiv.org/abs/1605.06061} {arXiv:1605.06061 [hep-ph]} \BibitemShut
  {NoStop}%
\bibitem [{\citenamefont {Das}\ \emph {et~al.}(2017)\citenamefont {Das},
  \citenamefont {Plumari}, \citenamefont {Chatterjee}, \citenamefont {Alam},
  \citenamefont {Scardina},\ and\ \citenamefont {Greco}}]{Das:2016cwd}%
  \BibitemOpen
  \bibfield  {author} {\bibinfo {author} {\bibfnamefont {S.~K.}\ \bibnamefont
  {Das}}, \bibinfo {author} {\bibfnamefont {S.}~\bibnamefont {Plumari}},
  \bibinfo {author} {\bibfnamefont {S.}~\bibnamefont {Chatterjee}}, \bibinfo
  {author} {\bibfnamefont {J.}~\bibnamefont {Alam}}, \bibinfo {author}
  {\bibfnamefont {F.}~\bibnamefont {Scardina}}, \ and\ \bibinfo {author}
  {\bibfnamefont {V.}~\bibnamefont {Greco}},\ }\href {\doibase
  10.1016/j.physletb.2017.02.046} {\bibfield  {journal} {\bibinfo  {journal}
  {Phys. Lett.}\ }\textbf {\bibinfo {volume} {B768}},\ \bibinfo {pages} {260}
  (\bibinfo {year} {2017})},\ \Eprint {http://arxiv.org/abs/1608.02231}
  {arXiv:1608.02231 [nucl-th]} \BibitemShut {NoStop}%
\bibitem [{\citenamefont {Dudal}\ and\ \citenamefont
  {Mertens}(2018)}]{Dudal:2018rki}%
  \BibitemOpen
  \bibfield  {author} {\bibinfo {author} {\bibfnamefont {D.}~\bibnamefont
  {Dudal}}\ and\ \bibinfo {author} {\bibfnamefont {T.~G.}\ \bibnamefont
  {Mertens}},\ }\href {\doibase 10.1103/PhysRevD.97.054035} {\bibfield
  {journal} {\bibinfo  {journal} {Phys. Rev.}\ }\textbf {\bibinfo {volume}
  {D97}},\ \bibinfo {pages} {054035} (\bibinfo {year} {2018})},\ \Eprint
  {http://arxiv.org/abs/1802.02805} {arXiv:1802.02805 [hep-th]} \BibitemShut
  {NoStop}%
\bibitem [{\citenamefont {Chatterjee}\ and\ \citenamefont
  {Bozek}(2018)}]{Chatterjee:2018lsx}%
  \BibitemOpen
  \bibfield  {author} {\bibinfo {author} {\bibfnamefont {S.}~\bibnamefont
  {Chatterjee}}\ and\ \bibinfo {author} {\bibfnamefont {P.}~\bibnamefont
  {Bozek}},\ }\href@noop {} {\  (\bibinfo {year} {2018})},\ \Eprint
  {http://arxiv.org/abs/1804.04893} {arXiv:1804.04893 [nucl-th]} \BibitemShut
  {NoStop}%
\bibitem [{\citenamefont {Rajagopal}\ and\ \citenamefont
  {Sadofyev}(2015)}]{Rajagopal:2015roa}%
  \BibitemOpen
  \bibfield  {author} {\bibinfo {author} {\bibfnamefont {K.}~\bibnamefont
  {Rajagopal}}\ and\ \bibinfo {author} {\bibfnamefont {A.~V.}\ \bibnamefont
  {Sadofyev}},\ }\href {\doibase 10.1007/JHEP10(2015)018} {\bibfield  {journal}
  {\bibinfo  {journal} {JHEP}\ }\textbf {\bibinfo {volume} {10}},\ \bibinfo
  {pages} {018} (\bibinfo {year} {2015})},\ \Eprint
  {http://arxiv.org/abs/1505.07379} {arXiv:1505.07379 [hep-th]} \BibitemShut
  {NoStop}%
\bibitem [{\citenamefont {Stephanov}\ and\ \citenamefont
  {Yee}(2016)}]{Stephanov:2015roa}%
  \BibitemOpen
  \bibfield  {author} {\bibinfo {author} {\bibfnamefont {M.~A.}\ \bibnamefont
  {Stephanov}}\ and\ \bibinfo {author} {\bibfnamefont {H.-U.}\ \bibnamefont
  {Yee}},\ }\href {\doibase 10.1103/PhysRevLett.116.122302} {\bibfield
  {journal} {\bibinfo  {journal} {Phys. Rev. Lett.}\ }\textbf {\bibinfo
  {volume} {116}},\ \bibinfo {pages} {122302} (\bibinfo {year} {2016})},\
  \Eprint {http://arxiv.org/abs/1508.02396} {arXiv:1508.02396 [hep-th]}
  \BibitemShut {NoStop}%
\bibitem [{\citenamefont {Sadofyev}\ and\ \citenamefont
  {Yin}(2016{\natexlab{b}})}]{Sadofyev:2015tmb}%
  \BibitemOpen
  \bibfield  {author} {\bibinfo {author} {\bibfnamefont {A.~V.}\ \bibnamefont
  {Sadofyev}}\ and\ \bibinfo {author} {\bibfnamefont {Y.}~\bibnamefont {Yin}},\
  }\href {\doibase 10.1103/PhysRevD.93.125026} {\bibfield  {journal} {\bibinfo
  {journal} {Phys. Rev.}\ }\textbf {\bibinfo {volume} {D93}},\ \bibinfo {pages}
  {125026} (\bibinfo {year} {2016}{\natexlab{b}})},\ \Eprint
  {http://arxiv.org/abs/1511.08794} {arXiv:1511.08794 [hep-th]} \BibitemShut
  {NoStop}%
\bibitem [{\citenamefont {Ozaki}\ \emph {et~al.}(2016)\citenamefont {Ozaki},
  \citenamefont {Itakura},\ and\ \citenamefont {Kuramoto}}]{Ozaki:2015sya}%
  \BibitemOpen
  \bibfield  {author} {\bibinfo {author} {\bibfnamefont {S.}~\bibnamefont
  {Ozaki}}, \bibinfo {author} {\bibfnamefont {K.}~\bibnamefont {Itakura}}, \
  and\ \bibinfo {author} {\bibfnamefont {Y.}~\bibnamefont {Kuramoto}},\ }\href
  {\doibase 10.1103/PhysRevD.94.074013} {\bibfield  {journal} {\bibinfo
  {journal} {Phys. Rev.}\ }\textbf {\bibinfo {volume} {D94}},\ \bibinfo {pages}
  {074013} (\bibinfo {year} {2016})},\ \Eprint
  {http://arxiv.org/abs/1509.06966} {arXiv:1509.06966 [hep-ph]} \BibitemShut
  {NoStop}%
\bibitem [{\citenamefont {Pineda}\ and\ \citenamefont
  {Soto}(1998)}]{Pineda:1997bj}%
  \BibitemOpen
  \bibfield  {author} {\bibinfo {author} {\bibfnamefont {A.}~\bibnamefont
  {Pineda}}\ and\ \bibinfo {author} {\bibfnamefont {J.}~\bibnamefont {Soto}},\
  }\href {\doibase 10.1016/S0920-5632(97)01102-X} {\bibfield  {journal}
  {\bibinfo  {journal} {Nucl. Phys. Proc. Suppl.}\ }\textbf {\bibinfo {volume}
  {64}},\ \bibinfo {pages} {428} (\bibinfo {year} {1998})},\ \Eprint
  {http://arxiv.org/abs/hep-ph/9707481} {arXiv:hep-ph/9707481 [hep-ph]}
  \BibitemShut {NoStop}%
\bibitem [{\citenamefont {Brambilla}\ \emph {et~al.}(2000)\citenamefont
  {Brambilla}, \citenamefont {Pineda}, \citenamefont {Soto},\ and\
  \citenamefont {Vairo}}]{Brambilla:1999xf}%
  \BibitemOpen
  \bibfield  {author} {\bibinfo {author} {\bibfnamefont {N.}~\bibnamefont
  {Brambilla}}, \bibinfo {author} {\bibfnamefont {A.}~\bibnamefont {Pineda}},
  \bibinfo {author} {\bibfnamefont {J.}~\bibnamefont {Soto}}, \ and\ \bibinfo
  {author} {\bibfnamefont {A.}~\bibnamefont {Vairo}},\ }\href {\doibase
  10.1016/S0550-3213(99)00693-8} {\bibfield  {journal} {\bibinfo  {journal}
  {Nucl. Phys.}\ }\textbf {\bibinfo {volume} {B566}},\ \bibinfo {pages} {275}
  (\bibinfo {year} {2000})},\ \Eprint {http://arxiv.org/abs/hep-ph/9907240}
  {arXiv:hep-ph/9907240 [hep-ph]} \BibitemShut {NoStop}%
\bibitem [{\citenamefont {Brambilla}\ \emph {et~al.}(2005)\citenamefont
  {Brambilla}, \citenamefont {Pineda}, \citenamefont {Soto},\ and\
  \citenamefont {Vairo}}]{Brambilla:2004jw}%
  \BibitemOpen
  \bibfield  {author} {\bibinfo {author} {\bibfnamefont {N.}~\bibnamefont
  {Brambilla}}, \bibinfo {author} {\bibfnamefont {A.}~\bibnamefont {Pineda}},
  \bibinfo {author} {\bibfnamefont {J.}~\bibnamefont {Soto}}, \ and\ \bibinfo
  {author} {\bibfnamefont {A.}~\bibnamefont {Vairo}},\ }\href {\doibase
  10.1103/RevModPhys.77.1423} {\bibfield  {journal} {\bibinfo  {journal} {Rev.
  Mod. Phys.}\ }\textbf {\bibinfo {volume} {77}},\ \bibinfo {pages} {1423}
  (\bibinfo {year} {2005})},\ \Eprint {http://arxiv.org/abs/hep-ph/0410047}
  {arXiv:hep-ph/0410047 [hep-ph]} \BibitemShut {NoStop}%
\bibitem [{\citenamefont {Pineda}(2012)}]{Pineda:2011dg}%
  \BibitemOpen
  \bibfield  {author} {\bibinfo {author} {\bibfnamefont {A.}~\bibnamefont
  {Pineda}},\ }\href {\doibase 10.1016/j.ppnp.2012.01.038} {\bibfield
  {journal} {\bibinfo  {journal} {Prog. Part. Nucl. Phys.}\ }\textbf {\bibinfo
  {volume} {67}},\ \bibinfo {pages} {735} (\bibinfo {year} {2012})},\ \Eprint
  {http://arxiv.org/abs/1111.0165} {arXiv:1111.0165 [hep-ph]} \BibitemShut
  {NoStop}%
\bibitem [{\citenamefont {Brambilla}\ \emph {et~al.}(2006)\citenamefont
  {Brambilla}, \citenamefont {Jia},\ and\ \citenamefont
  {Vairo}}]{Brambilla:2005zw}%
  \BibitemOpen
  \bibfield  {author} {\bibinfo {author} {\bibfnamefont {N.}~\bibnamefont
  {Brambilla}}, \bibinfo {author} {\bibfnamefont {Y.}~\bibnamefont {Jia}}, \
  and\ \bibinfo {author} {\bibfnamefont {A.}~\bibnamefont {Vairo}},\ }\href
  {\doibase 10.1103/PhysRevD.73.054005} {\bibfield  {journal} {\bibinfo
  {journal} {Phys. Rev.}\ }\textbf {\bibinfo {volume} {D73}},\ \bibinfo {pages}
  {054005} (\bibinfo {year} {2006})},\ \Eprint
  {http://arxiv.org/abs/hep-ph/0512369} {arXiv:hep-ph/0512369 [hep-ph]}
  \BibitemShut {NoStop}%
\bibitem [{\citenamefont {Brambilla}\ \emph {et~al.}(2012)\citenamefont
  {Brambilla}, \citenamefont {Pietrulewicz},\ and\ \citenamefont
  {Vairo}}]{Brambilla:2012be}%
  \BibitemOpen
  \bibfield  {author} {\bibinfo {author} {\bibfnamefont {N.}~\bibnamefont
  {Brambilla}}, \bibinfo {author} {\bibfnamefont {P.}~\bibnamefont
  {Pietrulewicz}}, \ and\ \bibinfo {author} {\bibfnamefont {A.}~\bibnamefont
  {Vairo}},\ }\href {\doibase 10.1103/PhysRevD.85.094005} {\bibfield  {journal}
  {\bibinfo  {journal} {Phys. Rev.}\ }\textbf {\bibinfo {volume} {D85}},\
  \bibinfo {pages} {094005} (\bibinfo {year} {2012})},\ \Eprint
  {http://arxiv.org/abs/1203.3020} {arXiv:1203.3020 [hep-ph]} \BibitemShut
  {NoStop}%
\bibitem [{\citenamefont {Pineda}\ and\ \citenamefont
  {Segovia}(2013)}]{Pineda:2013lta}%
  \BibitemOpen
  \bibfield  {author} {\bibinfo {author} {\bibfnamefont {A.}~\bibnamefont
  {Pineda}}\ and\ \bibinfo {author} {\bibfnamefont {J.}~\bibnamefont
  {Segovia}},\ }\href {\doibase 10.1103/PhysRevD.87.074024} {\bibfield
  {journal} {\bibinfo  {journal} {Phys. Rev.}\ }\textbf {\bibinfo {volume}
  {D87}},\ \bibinfo {pages} {074024} (\bibinfo {year} {2013})},\ \Eprint
  {http://arxiv.org/abs/1302.3528} {arXiv:1302.3528 [hep-ph]} \BibitemShut
  {NoStop}%
\bibitem [{\citenamefont {Faccioli}\ \emph {et~al.}(2011)\citenamefont
  {Faccioli}, \citenamefont {Lourenco}, \citenamefont {Seixas},\ and\
  \citenamefont {Wohri}}]{Faccioli:2011be}%
  \BibitemOpen
  \bibfield  {author} {\bibinfo {author} {\bibfnamefont {P.}~\bibnamefont
  {Faccioli}}, \bibinfo {author} {\bibfnamefont {C.}~\bibnamefont {Lourenco}},
  \bibinfo {author} {\bibfnamefont {J.}~\bibnamefont {Seixas}}, \ and\ \bibinfo
  {author} {\bibfnamefont {H.~K.}\ \bibnamefont {Wohri}},\ }\href {\doibase
  10.1103/PhysRevD.83.096001} {\bibfield  {journal} {\bibinfo  {journal} {Phys.
  Rev.}\ }\textbf {\bibinfo {volume} {D83}},\ \bibinfo {pages} {096001}
  (\bibinfo {year} {2011})},\ \Eprint {http://arxiv.org/abs/1103.4882}
  {arXiv:1103.4882 [hep-ph]} \BibitemShut {NoStop}%
\bibitem [{\citenamefont {Shao}\ and\ \citenamefont
  {Chao}(2014)}]{Shao:2012fs}%
  \BibitemOpen
  \bibfield  {author} {\bibinfo {author} {\bibfnamefont {H.-S.}\ \bibnamefont
  {Shao}}\ and\ \bibinfo {author} {\bibfnamefont {K.-T.}\ \bibnamefont
  {Chao}},\ }\href {\doibase 10.1103/PhysRevD.90.014002} {\bibfield  {journal}
  {\bibinfo  {journal} {Phys. Rev.}\ }\textbf {\bibinfo {volume} {D90}},\
  \bibinfo {pages} {014002} (\bibinfo {year} {2014})},\ \Eprint
  {http://arxiv.org/abs/1209.4610} {arXiv:1209.4610 [hep-ph]} \BibitemShut
  {NoStop}%
\bibitem [{\citenamefont {Shao}\ \emph {et~al.}(2014)\citenamefont {Shao},
  \citenamefont {Ma}, \citenamefont {Wang},\ and\ \citenamefont
  {Chao}}]{Shao:2014fca}%
  \BibitemOpen
  \bibfield  {author} {\bibinfo {author} {\bibfnamefont {H.-S.}\ \bibnamefont
  {Shao}}, \bibinfo {author} {\bibfnamefont {Y.-Q.}\ \bibnamefont {Ma}},
  \bibinfo {author} {\bibfnamefont {K.}~\bibnamefont {Wang}}, \ and\ \bibinfo
  {author} {\bibfnamefont {K.-T.}\ \bibnamefont {Chao}},\ }\href {\doibase
  10.1103/PhysRevLett.112.182003} {\bibfield  {journal} {\bibinfo  {journal}
  {Phys. Rev. Lett.}\ }\textbf {\bibinfo {volume} {112}},\ \bibinfo {pages}
  {182003} (\bibinfo {year} {2014})},\ \Eprint {http://arxiv.org/abs/1402.2913}
  {arXiv:1402.2913 [hep-ph]} \BibitemShut {NoStop}%
\bibitem [{\citenamefont {Bonati}\ \emph {et~al.}(2014)\citenamefont {Bonati},
  \citenamefont {D'Elia}, \citenamefont {Mariti}, \citenamefont {Mesiti},
  \citenamefont {Negro},\ and\ \citenamefont {Sanfilippo}}]{Bonati:2014ksa}%
  \BibitemOpen
  \bibfield  {author} {\bibinfo {author} {\bibfnamefont {C.}~\bibnamefont
  {Bonati}}, \bibinfo {author} {\bibfnamefont {M.}~\bibnamefont {D'Elia}},
  \bibinfo {author} {\bibfnamefont {M.}~\bibnamefont {Mariti}}, \bibinfo
  {author} {\bibfnamefont {M.}~\bibnamefont {Mesiti}}, \bibinfo {author}
  {\bibfnamefont {F.}~\bibnamefont {Negro}}, \ and\ \bibinfo {author}
  {\bibfnamefont {F.}~\bibnamefont {Sanfilippo}},\ }\href {\doibase
  10.1103/PhysRevD.89.114502} {\bibfield  {journal} {\bibinfo  {journal} {Phys.
  Rev.}\ }\textbf {\bibinfo {volume} {D89}},\ \bibinfo {pages} {114502}
  (\bibinfo {year} {2014})},\ \Eprint {http://arxiv.org/abs/1403.6094}
  {arXiv:1403.6094 [hep-lat]} \BibitemShut {NoStop}%
\bibitem [{\citenamefont {Steinbeiser}\ and\ \citenamefont
  {Segovia}(2017)}]{Steinbeisser:2017hke}%
  \BibitemOpen
  \bibfield  {author} {\bibinfo {author} {\bibfnamefont {S.}~\bibnamefont
  {Steinbeiser}}\ and\ \bibinfo {author} {\bibfnamefont {J.}~\bibnamefont
  {Segovia}},\ }\href {\doibase 10.1051/epjconf/201713706026} {\bibfield
  {journal} {\bibinfo  {journal} {EPJ Web Conf.}\ }\textbf {\bibinfo {volume}
  {137}},\ \bibinfo {pages} {06026} (\bibinfo {year} {2017})},\ \Eprint
  {http://arxiv.org/abs/1701.02513} {arXiv:1701.02513 [hep-ph]} \BibitemShut
  {NoStop}%
\bibitem [{\citenamefont {Segovia}\ and\ \citenamefont
  {Steinbeiser}(2018)}]{Segovia:2017nla}%
  \BibitemOpen
  \bibfield  {author} {\bibinfo {author} {\bibfnamefont {J.}~\bibnamefont
  {Segovia}}\ and\ \bibinfo {author} {\bibfnamefont {S.}~\bibnamefont
  {Steinbeiser}},\ }\href {\doibase 10.1088/1742-6596/1024/1/012016} {\bibfield
   {journal} {\bibinfo  {journal} {J. Phys. Conf. Ser.}\ }\textbf {\bibinfo
  {volume} {1024}},\ \bibinfo {pages} {012016} (\bibinfo {year} {2018})},\
  \Eprint {http://arxiv.org/abs/1708.08465} {arXiv:1708.08465 [hep-ph]}
  \BibitemShut {NoStop}%
\bibitem [{\citenamefont {Miransky}\ and\ \citenamefont
  {Shovkovy}(2002)}]{Miransky:2002rp}%
  \BibitemOpen
  \bibfield  {author} {\bibinfo {author} {\bibfnamefont {V.~A.}\ \bibnamefont
  {Miransky}}\ and\ \bibinfo {author} {\bibfnamefont {I.~A.}\ \bibnamefont
  {Shovkovy}},\ }\href {\doibase 10.1103/PhysRevD.66.045006} {\bibfield
  {journal} {\bibinfo  {journal} {Phys. Rev.}\ }\textbf {\bibinfo {volume}
  {D66}},\ \bibinfo {pages} {045006} (\bibinfo {year} {2002})},\ \Eprint
  {http://arxiv.org/abs/hep-ph/0205348} {arXiv:hep-ph/0205348 [hep-ph]}
  \BibitemShut {NoStop}%
\bibitem [{\citenamefont {Andreichikov}\ \emph {et~al.}(2013)\citenamefont
  {Andreichikov}, \citenamefont {Orlovsky},\ and\ \citenamefont
  {Simonov}}]{Andreichikov:2012xe}%
  \BibitemOpen
  \bibfield  {author} {\bibinfo {author} {\bibfnamefont {M.~A.}\ \bibnamefont
  {Andreichikov}}, \bibinfo {author} {\bibfnamefont {V.~D.}\ \bibnamefont
  {Orlovsky}}, \ and\ \bibinfo {author} {\bibfnamefont {{\relax Yu}.~A.}\
  \bibnamefont {Simonov}},\ }\href {\doibase 10.1103/PhysRevLett.110.162002}
  {\bibfield  {journal} {\bibinfo  {journal} {Phys. Rev. Lett.}\ }\textbf
  {\bibinfo {volume} {110}},\ \bibinfo {pages} {162002} (\bibinfo {year}
  {2013})},\ \Eprint {http://arxiv.org/abs/1211.6568} {arXiv:1211.6568
  [hep-ph]} \BibitemShut {NoStop}%
\bibitem [{\citenamefont {Chernodub}(2014)}]{Chernodub:2014uua}%
  \BibitemOpen
  \bibfield  {author} {\bibinfo {author} {\bibfnamefont {M.~N.}\ \bibnamefont
  {Chernodub}},\ }\href {\doibase 10.1142/S0217732314501624} {\bibfield
  {journal} {\bibinfo  {journal} {Mod. Phys. Lett.}\ }\textbf {\bibinfo
  {volume} {A29}},\ \bibinfo {pages} {1450162} (\bibinfo {year}
  {2014})}\BibitemShut {NoStop}%
\bibitem [{\citenamefont {Rougemont}\ \emph {et~al.}(2015)\citenamefont
  {Rougemont}, \citenamefont {Critelli},\ and\ \citenamefont
  {Noronha}}]{Rougemont:2014efa}%
  \BibitemOpen
  \bibfield  {author} {\bibinfo {author} {\bibfnamefont {R.}~\bibnamefont
  {Rougemont}}, \bibinfo {author} {\bibfnamefont {R.}~\bibnamefont {Critelli}},
  \ and\ \bibinfo {author} {\bibfnamefont {J.}~\bibnamefont {Noronha}},\ }\href
  {\doibase 10.1103/PhysRevD.91.066001} {\bibfield  {journal} {\bibinfo
  {journal} {Phys. Rev.}\ }\textbf {\bibinfo {volume} {D91}},\ \bibinfo {pages}
  {066001} (\bibinfo {year} {2015})},\ \Eprint {http://arxiv.org/abs/1409.0556}
  {arXiv:1409.0556 [hep-th]} \BibitemShut {NoStop}%
\bibitem [{\citenamefont {Simonov}\ and\ \citenamefont
  {Trusov}(2015)}]{Simonov:2015yka}%
  \BibitemOpen
  \bibfield  {author} {\bibinfo {author} {\bibfnamefont {{\relax Yu}.~A.}\
  \bibnamefont {Simonov}}\ and\ \bibinfo {author} {\bibfnamefont {M.~A.}\
  \bibnamefont {Trusov}},\ }\href {\doibase 10.1016/j.physletb.2015.05.032}
  {\bibfield  {journal} {\bibinfo  {journal} {Phys. Lett.}\ }\textbf {\bibinfo
  {volume} {B747}},\ \bibinfo {pages} {48} (\bibinfo {year} {2015})},\ \Eprint
  {http://arxiv.org/abs/1503.08531} {arXiv:1503.08531 [hep-ph]} \BibitemShut
  {NoStop}%
\bibitem [{\citenamefont {Dudal}\ and\ \citenamefont
  {Mahapatra}(2017)}]{Dudal:2016joz}%
  \BibitemOpen
  \bibfield  {author} {\bibinfo {author} {\bibfnamefont {D.}~\bibnamefont
  {Dudal}}\ and\ \bibinfo {author} {\bibfnamefont {S.}~\bibnamefont
  {Mahapatra}},\ }\href {\doibase 10.1007/JHEP04(2017)031} {\bibfield
  {journal} {\bibinfo  {journal} {JHEP}\ }\textbf {\bibinfo {volume} {04}},\
  \bibinfo {pages} {031} (\bibinfo {year} {2017})},\ \Eprint
  {http://arxiv.org/abs/1612.06248} {arXiv:1612.06248 [hep-th]} \BibitemShut
  {NoStop}%
\bibitem [{\citenamefont {Hasan}\ \emph {et~al.}(2017)\citenamefont {Hasan},
  \citenamefont {Chatterjee},\ and\ \citenamefont {Patra}}]{Hasan:2017fmf}%
  \BibitemOpen
  \bibfield  {author} {\bibinfo {author} {\bibfnamefont {M.}~\bibnamefont
  {Hasan}}, \bibinfo {author} {\bibfnamefont {B.}~\bibnamefont {Chatterjee}}, \
  and\ \bibinfo {author} {\bibfnamefont {B.~K.}\ \bibnamefont {Patra}},\ }\href
  {\doibase 10.1140/epjc/s10052-017-5346-z} {\bibfield  {journal} {\bibinfo
  {journal} {Eur. Phys. J.}\ }\textbf {\bibinfo {volume} {C77}},\ \bibinfo
  {pages} {767} (\bibinfo {year} {2017})},\ \Eprint
  {http://arxiv.org/abs/1703.10508} {arXiv:1703.10508 [hep-ph]} \BibitemShut
  {NoStop}%
\bibitem [{\citenamefont {Singh}\ \emph {et~al.}(2017)\citenamefont {Singh},
  \citenamefont {Thakur},\ and\ \citenamefont {Mishra}}]{Singh:2017nfa}%
  \BibitemOpen
  \bibfield  {author} {\bibinfo {author} {\bibfnamefont {B.}~\bibnamefont
  {Singh}}, \bibinfo {author} {\bibfnamefont {L.}~\bibnamefont {Thakur}}, \
  and\ \bibinfo {author} {\bibfnamefont {H.}~\bibnamefont {Mishra}},\
  }\href@noop {} {\  (\bibinfo {year} {2017})},\ \Eprint
  {http://arxiv.org/abs/1711.03071} {arXiv:1711.03071 [hep-ph]} \BibitemShut
  {NoStop}%
\bibitem [{\citenamefont {Bonati}\ \emph {et~al.}(2016)\citenamefont {Bonati},
  \citenamefont {D'Elia}, \citenamefont {Mariti}, \citenamefont {Mesiti},
  \citenamefont {Negro}, \citenamefont {Rucci},\ and\ \citenamefont
  {Sanfilippo}}]{Bonati:2016kxj}%
  \BibitemOpen
  \bibfield  {author} {\bibinfo {author} {\bibfnamefont {C.}~\bibnamefont
  {Bonati}}, \bibinfo {author} {\bibfnamefont {M.}~\bibnamefont {D'Elia}},
  \bibinfo {author} {\bibfnamefont {M.}~\bibnamefont {Mariti}}, \bibinfo
  {author} {\bibfnamefont {M.}~\bibnamefont {Mesiti}}, \bibinfo {author}
  {\bibfnamefont {F.}~\bibnamefont {Negro}}, \bibinfo {author} {\bibfnamefont
  {A.}~\bibnamefont {Rucci}}, \ and\ \bibinfo {author} {\bibfnamefont
  {F.}~\bibnamefont {Sanfilippo}},\ }\href {\doibase
  10.1103/PhysRevD.94.094007} {\bibfield  {journal} {\bibinfo  {journal} {Phys.
  Rev.}\ }\textbf {\bibinfo {volume} {D94}},\ \bibinfo {pages} {094007}
  (\bibinfo {year} {2016})},\ \Eprint {http://arxiv.org/abs/1607.08160}
  {arXiv:1607.08160 [hep-lat]} \BibitemShut {NoStop}%
\bibitem [{\citenamefont {Bonati}\ \emph {et~al.}(2017)\citenamefont {Bonati},
  \citenamefont {D'Elia}, \citenamefont {Mariti}, \citenamefont {Mesiti},
  \citenamefont {Negro}, \citenamefont {Rucci},\ and\ \citenamefont
  {Sanfilippo}}]{Bonati:2017uvz}%
  \BibitemOpen
  \bibfield  {author} {\bibinfo {author} {\bibfnamefont {C.}~\bibnamefont
  {Bonati}}, \bibinfo {author} {\bibfnamefont {M.}~\bibnamefont {D'Elia}},
  \bibinfo {author} {\bibfnamefont {M.}~\bibnamefont {Mariti}}, \bibinfo
  {author} {\bibfnamefont {M.}~\bibnamefont {Mesiti}}, \bibinfo {author}
  {\bibfnamefont {F.}~\bibnamefont {Negro}}, \bibinfo {author} {\bibfnamefont
  {A.}~\bibnamefont {Rucci}}, \ and\ \bibinfo {author} {\bibfnamefont
  {F.}~\bibnamefont {Sanfilippo}},\ }\href {\doibase
  10.1103/PhysRevD.95.074515} {\bibfield  {journal} {\bibinfo  {journal} {Phys.
  Rev.}\ }\textbf {\bibinfo {volume} {D95}},\ \bibinfo {pages} {074515}
  (\bibinfo {year} {2017})},\ \Eprint {http://arxiv.org/abs/1703.00842}
  {arXiv:1703.00842 [hep-lat]} \BibitemShut {NoStop}%
\end{thebibliography}%

\end{document}